\documentclass[12pt]{article}
\usepackage{bm}
\usepackage{a4wide,graphicx}
\usepackage{cite}
\usepackage{amssymb}
\usepackage{amsmath}
\usepackage{color}

\newcommand{\vecsig}{\vec\sigma}
\newcommand{\vecq}{\vec q}

\begin{document}


\title{Anti-strange meson-baryon interaction \\
 in hot and dense nuclear matter}

\author{D.~Cabrera$^{1,2}$, L.~Tol\'os$^{2,3}$, J.~Aichelin$^4$ and E.~Bratkovskaya$^{1,2}$\\
$^1$ Institut f\"ur Theoretische Physik, Johann Wolfgang\\
Goethe-Universit\"at Frankfurt am Main, 60438 Frankfurt am Main, Germany\\
$^2$ Frankfurt Institute for Advanced Studies (FIAS),\\
60438 Frankfurt am Main, Germany\\
$^3$ Institut de Ci\`encies de l'Espai (IEEC/CSIC), Campus Universitat\\
Aut\`onoma de Barcelona, Facultat de Ci\`encies, Torre C5,\\
E-08193 Bellaterra, Spain\\
$^4$ Subatech, UMR 6457, IN2P3/CNRS, Universit\'e de Nantes,\\
\'Ecole des Mines de Nantes, 4 rue Alfred Kastler, 44307 Nantes cedex 3, France
}

\date{\today}

\maketitle
\begin{abstract}
We present a study of in-medium cross sections and (off-shell) transition rates
for the most relevant binary reactions for strange pseudoscalar meson production
close to threshold in heavy-ion collisions at FAIR energies.
Our results rely on a chiral unitary approach in coupled channels which
incorporates the $s$- and $p$-waves of the kaon-nucleon interaction. The formalism,
which is modified in the hot and dense medium to
account for Pauli blocking effects, mean-field binding on
baryons, and pion and kaon self-energies, has been improved to implement
unitarization and self-consistency for both the $s$- and $p$-wave interactions
at finite temperature and density. This gives access to in-medium amplitudes
in several elastic and inelastic coupled channels with strangeness content $S=-1$.
The obtained total cross sections mostly reflect the fate of the $\Lambda(1405)$
resonance, which melts in the nuclear environment, whereas the off-shell
transition probabilities are also sensitive to the in-medium properties of the
hyperons excited in the $p$-wave amplitudes [$\Lambda$, $\Sigma$ and $\Sigma^*(1385)$].
The single-particle  potentials of these hyperons at finite momentum, density and
temperature are also discussed in connection with the pertinent scattering amplitudes.
Our results are the basis for future implementations in microscopic transport
approaches accounting for off-shell dynamics of strangeness production in nucleus-nucleus collisions.

\end{abstract}
\vskip 0.5 cm



\section{Introduction}  
\label{sec:intro}

Strongly interacting matter under extreme conditions of temperature and
density has been a matter of interest over the last decades, related to
the understanding of the strong interaction  \cite{Rapp:1999ej} as well
as the analysis of compact stars. In particular, strange pseudoscalar
mesons  in matter have been throughly investigated in exotic atoms
\cite{Friedman:2007zza}, 
  heavy-ion collisions (HICs)
\cite{Aichelin:1986ss,Ko1,Cass:1999,Fuchs:2005zg,FOPI,KaoS,Forster:2007qk,Hartnack:2011cn,Zinyuk:2014zor}
and neutron stars \cite{Kaplan:1986yq}.

The phenomenology of kaonic atoms \cite{Friedman:2007zza} requires an
attractive potential for $\bar K$ mesons  whereas the $\bar K N$
scattering amplitude in vacuum is repulsive at low energies
due to the presence of the $\Lambda(1405)$ resonance below the $\bar K N$ threshold. Indeed, this resonance has been throughly analyzed in photon induced reactions by the CLAS collaboration \cite{Moriya:2013eb} and in proton-proton reactions by the ANKE experiment \cite{Zychor:2007gf}, and more recently by the HADES at GSI \cite{Agakishiev:2012xk}.
The onset
of an attractive $\bar K N$ interaction at low densities is a
consequence of an upper shift of the $\Lambda(1405)$ resonance induced
by Pauli blocking on the intermediate nucleon states
\cite{Koch:1994mj,Waas:1996xh,Waas:1996fy,Lutz:1997wt}. Additional
medium effects such as the self-energy of mesons in related coupled
channels and the binding of hyperons in the nuclear environment bring a
smoothened $\Lambda (1405)$ back to  its vacuum position
\cite{Ramos:1999ku}, while keeping the attractive character of the
$\bar K N$ interaction in matter.

Unitarized chiral coupled-channel approaches
\cite{Waas:1996xh,Waas:1996fy,Lutz:1997wt,Jido:2003cb,Weise:2008aj}
with a self-consistent evaluation of the kaon self-energy
\cite{Lutz:1997wt,Ramos:1999ku,Tolos:2000fj,Tolos:2006ny,Lutz:2007bh}
have proven to be very successful in describing the $\bar K$ meson
interaction in matter. An attractive potential of about 40-60~MeV at
normal nuclear matter density is obtained when self-consistency is
implemented, rather shallow as compared to relativistic mean-field
calculations \cite{Schaffner:1996kv} or phenomenological analysis of
kaonic atom data with density dependent potentials including
non-linearities  \cite{Friedman:2007zza,Cieply:2011fy,Friedman:2012qy}.
Yet, this shallow potential is able to reproduce the data from kaonic
atoms \cite{Hirenzaki:2000da,Baca:2000ic}.

The $\bar K$ meson interaction with nucleons has also been addressed
recently in connection to the possible formation of deeply bound kaonic
states after the prediction of narrow strongly bound states in few-body
systems \cite{Akaishi:2002bg,dote04,Akaishi:2005sn}. This analysis was
strongly criticized \cite{Oset:2005sn} due to the unrealistic treatment
of the $\bar K N$ interaction. Recent improved calculations using
different few-body methods with diverse  $\bar KN$ input
\cite{Shevchenko:2006xy,Shevchenko:2007zz,Ikeda:2007nz,Revai:2014twa,Ikeda:2008ub,Ikeda:2010tk,Dote:2008in,Dote:2008hw,Barnea:2012qa,Bayar:2012hn}  predict few-nucleon kaonic states with large widths, although the predicted binding energies and widths differ substantially from one model to the other. Thus, the experimental quest for such deeply bound kaonic states is an active field of research \cite{FINUDA,OBELIX,DISTO,HADES,LEPS,E15,E27}, that will allow to further constrain the $\bar K N$ interaction in the near future.

Moreover, the in-medium modification of kaon/antikaon properties has been
explored experimentally  close to threshold in heavy-ion collisions
at SIS energies \cite{FOPI,KaoS,Forster:2007qk}. With the help of
microscopic transport models
\cite{Aichelin:1986ss,Ko1,Cass:1999,Aichelin,Fuchs:2005zg,Cassing:2003vz,Hartnack:2011cn}
the creation and transport of kaons/antikaons has been studied revealing
a complicated multiple interaction scenario of the strange particles with
hadronic matter whose consequences show up in the measured
spectra and kaon flow characteristics. The strangeness production in heavy-ion collisions is very
different from that in elementary interactions as the excitation
functions for kaons and antikaons  show. The comparison of transport model calculations with
experimental results (such as production cross sections, energy
and polar angular distributions, azimuthal anisotropy coefficients
$v_1, v_2$ etc.) indicate that in matter the kaons are affected by
a shallow repulsive potential whereas the antikaon dynamics are influenced by a much
stronger attractive potential.

For kaons the spectral function is very narrow and therefore it
behaves almost as a good quasi-particle. For antikaons the situation is
much more uncertain. This is due to three reasons: a) They have a
broad spectral function due to strong interactions with the
baryons. b) The simple $t\, \rho$ approximation for the antikaon
optical potential does not work in the $I=0$ channel since this scattering amplitude is dominated by the
$\Lambda(1405)$ resonance and is repulsive in vacuum. c) The measured
excitation function of the antikaon yield close to  threshold
energies confirms that the $\pi Y \to {\bar K}N$ reaction is the dominant channel for
 antikaon production in heavy-ion collisions
\cite{Aichelin:1986ss} since the hyperons are more abundantly produced together with kaons.
This cross section is expected to be
substantially modified in the hot and dense medium. For all
these reasons it is very important to incorporate  a self-consistent treatment of the $\bar K$ self
energy and the $\bar K$ scattering amplitudes in transport
calculations.

The first transport calculations for antikaon observables in
nuclear matter were performed assuming that the finite width
of the antikaon spectral function might be neglected
\cite{Aichelin:1986ss,Cass:1999,Aichelin,Hartnack:2002xc}. These calculations revealed the strangeness exchange reaction as the
dominant production channel and the existence of an attractive
antikaon optical potential. Some years later antikaon production was
studied using off-shell dynamics with in-medium spectral functions
in the  Hadron-String-Dynamics (HSD) transport model
\cite{Cassing:2003vz} employing the J\"ulich meson-exchange model
\cite{Tolos:2000fj,Tolos:2002ud} as  the effective $\bar KN$
interaction in matter. Multiplicity ratios involving strange mesons
coming from heavy-ion collisions data were analyzed in
\cite{Tolos:2003qj}.

During the last decade several conclusions on the production mechanisms
for strangeness have been achieved by the analysis of experimental data
 in conjunction with microscopic transport approaches,
i.e. the production mechanisms of strangeness, the different freeze-out
conditions exhibited by $K^+$ and $K^-$ mesons and the use of $K^+$ as
a probe of the nuclear matter equation of state at high baryon densities.
Still, the analysis of all experimental antikaon observables has not allowed so far
for a consensus on the antikaon cross sections and optical potential
(cf. the recent review \cite{Hartnack:2011cn}).
For example, recent experimental data on the $v_1, v_2$ flow  of strange
mesons \cite{Zinyuk:2014zor} show a sensitivity to the
details of the in-medium meson-baryon interaction, leaving room for a
more elaborate description within hadronic models.

A model for the $\bar K N$ interaction at finite density and zero
temperature has been recently developed within a chiral unitarity
approach in coupled channels by incorporating the $s$- and $p$- waves
of the kaon-nucleon interaction in a self-consistent manner
\cite{Tolos:2006ny}. Finite temperature effects have been also
implemented \cite{Tolos:2008di}, although only a full self-consistent
solution for $s$- wave effective $\bar KN$ interaction was reached as
the $p$-wave contribution was treated by means of hyperon-nucleon
insertions. In this work we aim at improving on the chiral effective
scheme in dense matter developed in
Refs.~\cite{Tolos:2006ny,Tolos:2008di} as we incorporate the full
self-consistency in $s$- and $p$-waves at finite density and
temperature. In this way, we are able to generate in-medium
meson-baryon cross sections (amplitudes) at finite temperature as well
as to determine the single-particle properties of hyperons, such as
$\Lambda(1115)$, $\Sigma(1195)$ and $\Sigma^*(1385)$, at finite
momentum, density and temperature. These results will be used to
analyze the antikaon and hyperon production near threshold in HICs  in
a subsequent publication \cite{preparation}.

This paper is organized as follows. In Sec.~\ref{sec:model} we present
the improved model for the $S=-1$ meson-baryon amplitudes in hot
nuclear matter. In Sec.~\ref{sec:amplitudes} the $S=-1$ in-medium
amplitudes and the single-particle properties of the $\Lambda$,
$\Sigma$ and $\Sigma^*(1385)$ at finite density, temperature and
momentum are studied, whereas in Sec.~\ref{sec:cross-sec} we show the
results for the in-medium cross sections and transition amplitudes. We
draw our summary, conclusions and outlook in Sec.~\ref{sec:Conclusion}.

\section{Chiral unitarized model for $S=-1$  meson-baryon amplitudes in hot nuclear matter}     
\label{sec:model}

In the present work we build upon the recent results of Refs.~\cite{Tolos:2006ny,Tolos:2008di}, where the properties of strange mesons in nuclear matter at finite temperature were studied within a self-consistent coupled-channel approach based on the $SU(3)$ meson-baryon chiral Lagrangian.

In Ref.~\cite{Jido:2002zk} the $p$-wave amplitude in vacuum was added to the $s$-wave contribution coming from the Weinberg-Tomozawa term \cite{Oset:1997it}. The $p$-wave scattering was generated by the pole terms of the octet $\Lambda(1115)$, $\Sigma(1195)$ and the decuplet $\Sigma^*(1385)$ in $s$-channel exchange \cite{Jido:2002zk}. A full self-consistent treatment of the in-medium interaction at zero temperature in $s$- and $p$-waves was performed in a later work \cite{Tolos:2006ny}. In addition, nuclear short range correlations were incorporated in the $p$-wave amplitudes in line with the mechanisms that drive the nucleon-nucleon and nucleon-hyperon interactions in Ref.~\cite{Tolos:2006ny}, thus improving the formalism developed in \cite{Ramos:1999ku}.

The effect of finite temperature was taken into account  in Ref.~\cite{Tolos:2008di} by recalculating all the relevant meson-baryon propagators and self-energies within the Imaginary Time (Matsubara) Formalism, thus extending the applicability of the model to the experimental conditions of intermediate energy heavy-ion collisions (FAIR). Still, the $p$-wave self-energy of kaons and antikaons was calculated at the level of single hyperon-hole insertions and not within the present unitarized and self-consistent scheme. Thus, although we were able to obtain the $p$-wave self-energy, which was evaluated in terms of finite-temperature hyperon-hole Lindhard functions including baryonic mean-field potentials, a drawback of this calculation was that the in-medium $p$-wave amplitudes for $\bar K N \to \bar K N$ and related (off-diagonal) coupled channels were not accessible at finite temperature.

With the focus on the implementation of in-medium hadronic scattering amplitudes in microscopic transport simulations, we have improved our previous calculations in \cite{Tolos:2006ny} by adding the unitarization of the $\bar K N$ $p$-wave interaction and keeping the finite temperature formalism of \cite{Tolos:2008di} for the scattering amplitudes and the meson self-energies. This improvement not only generalizes the results of \cite{Tolos:2006ny} to hot and dense matter, but additionally gives access to full off-shell in-medium scattering amplitudes in the $SU(3)$ set of coupled channels.

Moreover, the improved model renders an additional output, namely the in-medium single-particle properties of the hyperons exchanged in the $p$-wave amplitudes, which are consistently generated within the same approach. Previous results in cold nuclear matter were advanced in \cite{Tolos:2006ny} for the mass shift and width of these states at normal matter density, $\rho_0=0.17 {\rm fm^{-3}}$. We generalize and extend those results by providing the density, temperature and momentum dependent single-particle potentials for the $\Lambda(1115)$, $\Sigma(1195)$ and $\Sigma^*(1385)$, which we obtain by analyzing the poles in the scattering amplitudes, cf.~Sec.~\ref{sec:amplitudes}.

The dynamics of strange meson-baryon scattering as can be extracted from our scattering amplitudes is best implemented within transport models in terms of in-medium cross sections or else as off-shell reaction rates when the propagation of unstable particles is taken into account \cite{Hartnack:2011cn}. We explore both scenarios and for the first time we calculate in our model the total cross section of several $\bar K N$ two-body reactions at finite temperature and nuclear density as well as the off-shell transition probabilities for several processes which play a key role in accessing the near sub-threshold region in anti-kaon production dynamics (cf.~Sec.~\ref{sec:cross-sec}).

\subsection{$S=-1$ meson-baryon amplitudes in vacuum}

The extensive details of the formalism for $\bar K N$ scattering and related channels in meson-baryon Chiral Perturbation Theory can be found in \cite{Oset:1997it,Oller:2000fj,Jido:2002zk,Jido:2003cb,Garcia-Recio:2003ks,Hyodo:2002pk,Borasoy:2005ie,Oller:2006jw,Borasoy:2006sr}. Here we provide a brief summary of the leading order $s$- and $p$-wave scattering amplitudes in vacuum, and the unitarization in coupled channels.

The lowest order chiral Lagrangian which couples the octet of light pseudoscalar
mesons to the octet of $1/2^+$ baryons is given by
\begin{eqnarray}
   {\cal L}_1^{(B)} &=& \langle \bar{B} i \gamma^{\mu} \nabla_{\mu} B
    \rangle  - M \langle \bar{B} B\rangle  \nonumber \\
    &&  + \frac{1}{2} D \left\langle \bar{B} \gamma^{\mu} \gamma_5 \left\{
     u_{\mu}, B \right\} \right\rangle + \frac{1}{2} F \left\langle \bar{B}
     \gamma^{\mu} \gamma_5 \left[u_{\mu}, B\right] \right\rangle  \ ,
    \label{chiralLag}
\end{eqnarray}
where the symbol $\langle \, \rangle$ denotes the trace of $SU(3)$ flavor
matrices, $M$ is the baryon mass and
$\nabla_{\mu}$ denotes the covariant derivative coupling the baryon fields to the pseudoscalar meson vector current $\Gamma_{\mu}$,
\begin{eqnarray}
\label{eq:defs1}
  \nabla_{\mu} B &=& \partial_{\mu} B + [\Gamma_{\mu}, B] \nonumber \ ,\\
  \Gamma_{\mu} &=& \frac{1}{2} (u^\dagger \partial_{\mu} u + u\, \partial_{\mu}
      u^\dagger) \ .
\end{eqnarray}
The pseudoscalar (Goldstone) bosons are introduced within the non-linear realization of chiral symmetry in exponential parameterization, $U = u^2 = {\rm exp} (i \sqrt{2} \Phi / f)$, and $f$ is the meson decay constant. The two last terms in Eq.~(\ref{chiralLag}) contain the coupling of the baryon fields to the meson axial vector current $u_{\mu}$, with
\begin{equation}
\label{eq:defs2}
  u_{\mu} = i u ^\dagger \partial_{\mu} U u^\dagger = i \left( u^\dagger \partial_{\mu} u
  - u \partial_{\mu} u^\dagger \right) .
\end{equation}
We note that in the $SU(2)$ sector only the sum $D+F$ is relevant and corresponds to the nucleon axial vector coupling. The $\pi NN$ interaction strength relates to the former via the Goldberger-Treiman relation, $g_{\pi NN}/2M_N = (D+F)/2f$.
The $SU(3)$ meson and baryon field matrices are standard in notation and given by
\begin{equation}
\Phi =
\left(
\begin{array}{ccc}
\frac{1}{\sqrt{2}} \pi^0 + \frac{1}{\sqrt{6}} \eta & \pi^+ & K^+ \\
\pi^- & - \frac{1}{\sqrt{2}} \pi^0 + \frac{1}{\sqrt{6}} \eta & K^0 \\
K^- & \bar{K}^0 & - \frac{2}{\sqrt{6}} \eta
\end{array}
\right) \ ,
\end{equation}
\begin{equation}
B =
\left(
\begin{array}{ccc}
\frac{1}{\sqrt{2}} \Sigma^0 + \frac{1}{\sqrt{6}} \Lambda &
\Sigma^+ & p \\
\Sigma^- & - \frac{1}{\sqrt{2}} \Sigma^0 + \frac{1}{\sqrt{6}} \Lambda & n \\
\Xi^- & \Xi^0 & - \frac{2}{\sqrt{6}} \Lambda
\end{array}
\right) \ .
\end{equation}

Let us focus first on the $s$-wave meson-baryon interaction. Keeping at the level of two meson fields, the covariant derivative term in
Eq.~(\ref{chiralLag}) provides the following interaction Lagrangian,
\begin{equation}
   {\cal L}_1^{(B)} \doteq \left\langle \bar{B} i \gamma^{\mu} \frac{1}{4 f^2}
   [(\Phi\, \partial_{\mu} \Phi - \partial_{\mu} \Phi \Phi) B
   - B (\Phi\, \partial_{\mu} \Phi - \partial_{\mu} \Phi \Phi)]
   \right\rangle \ , \label{lowest}
\end{equation}
from which one can derive the meson-baryon (tree-level) transition amplitudes as
\begin{equation}
   V_{ij} = - C_{ij} {1 \over 4 f^2} \bar{u}(p^\prime) \gamma^\mu u(p)
   (k_\mu + k^\prime_\mu)  \ ,\label{fourpoint}
\end{equation}
where $k$, $k^\prime$ ($p$, $p^\prime$) are the initial and final
meson (baryon) momenta, respectively, and the coefficients $C_{ij}$ 
($i$, $j$ indicate the particular meson-baryon channel) form a
symmetric matrix and can be found explicitly in
\cite{Oset:1997it}.
For low-energy scattering (i.e. neglecting corrections of order $p/M$) the following expression for the $s$-wave scattering amplitude is obtained,
\begin{eqnarray}
V_{i j}^s &=& - C_{i j} \, \frac{1}{4 f^2} \, (2 \, \sqrt{s}-M_{B_i}-M_{B_j})
\left( \frac{M_{B_i}+E_i}{2 \, M_{B_i}} \right)^{1/2} \, \left( \frac{M_{B_j}+E_j}{2 \, M_{B_j}} \right)^{1/2} \nonumber \\
&\simeq& - C_{i j} \, {1 \over 4 f^2} (k^0_i + k^0_j)
\ ,
\label{swa}
\end{eqnarray}
where $\sqrt{s}$ is the center-of-mass (c.m.) energy, $M_{B_{i(j)}}$ and $E_{i(j)}$ are the mass and energy of the baryon in the $i(j)$ channel, respectively, and the second equation is satisfied to a good approximation for practical purposes. Note that in the previous expressions the spin structure is omitted for simplicity of notation and a $\chi^\dagger_s \dots \chi_r$ spinor product has to be understood.
The meson decay constant $f$
is taken as an average value $f=1.123 f_\pi$ \cite{Oset:2001cn}, as is customary in meson-baryon studies within the strangeness -$1$ sector.
The channels included in our study are $K^- p$, $\bar{K}^0n$, $\pi^0
\Lambda$, $\pi^0 \Sigma^0$, $\eta \Lambda$, $\eta \Sigma^0$, $\pi^+
\Sigma^-$, $\pi^- \Sigma^+$, $K^+ \Xi^-$, $K^0 \Xi^0$.

Unitarization along the right-hand cut of the leading order (tree-level) amplitudes in a coupled-channel approach has been thoroughly established as the method to extend the applicability of the effective theory to higher energies and, in particular, to account for the presence of resonant states, such as the $s$-wave $\Lambda(1405)$. Formally, the unitarized solution is obtained by iteration of the leading order amplitude in a Bethe-Salpeter equation in coupled channels (in matrix notation),
\begin{equation}
\label{eq:BS-matrix}
T = V+\overline{VGT} \ ,
\end{equation}
where $V$ is the $s$-wave potential discussed above and the line indicates the phase-space integral over intermediate meson-baryon states in the $VGT$ term. The set of coupled integral equations involved in Eq.~(\ref{eq:BS-matrix}) is notably simplified within the chiral effective theory since both the potential $V$ and the resummed amplitude $T$ can be factorized on-shell, and thus the solution proceeds by algebraic inversion, $T=[1-VG]^{-1} V$.
For $s$-wave amplitudes it has been shown that the off-shell parts in the integral term of the equation lead to structures that can be renormalized by higher-order counterterms and are effectively accounted for by using physical masses and coupling constants \cite{Oset:1997it}.  A more general proof of the on-shell factorization in absence of a left-hand cut was given in \cite{Oller:1998zr,Oller:2000fj} based on the $N/D$ method and dispersion relations. The quantity $G$ is a diagonal matrix accounting for the meson-baryon loop function,
\begin{equation}
\label{G_vacuum}
G_l (\sqrt{s}) = {\rm i} \,
\int \frac{d^4q}{(2\, \pi)^4} \,
\frac{M_{l}}{E_l(\vec{P}-\vec{q}\,)} \,
\frac{1}{\sqrt{s} - q_0 - E_l(\vec{P}-\vec{q}\,) + {\rm i} \varepsilon} \,
\frac{1}{q_0^2 - \vec{q}\,^2 - m_l^2 + {\rm i} \varepsilon} \ \,
\end{equation}
with $(P^0,\vec{P})$ being the total four-momentum of the meson-baryon pair and $s=(P^0)^2-\vec{P}\,^2$. Note that we work with a non-relativistic reduction of baryon propagators (leading order in $M_B^{-1}$) in consistency with the approximations done in Eq.~(\ref{swa}), and therefore we neglect contributions from negative-energy poles (we keep, however, full relativistic kinematics for the baryon dispersion relation).
The loop function is divergent and needs to be regularized. This can be done by adopting either a cutoff method or dimensional regularization. Both schemes provide equivalent results as the pertinent regularization parameters (cut-off momentum, $q_{\rm max}$, and subtraction constant, $a_{MB}$) can be related at a given energy scale \cite{Oller:1997ng}. For practical purposes the cutoff method is more convenient and transparent when dealing with particles in the medium. Within this method, and taking advantage of Lorentz invariance to calculate in the c.m. frame, the loop function reads
\begin{eqnarray}
\hspace{-0.5cm}G_{l}(\sqrt{s})&=& i \, \int \frac{d^4 q}{(2
\pi)^4} \, \frac{M_l}{E_l (-\vec{q}\,)} \, \frac{1}{\sqrt{s} - q^0 - E_l
(-\vec{q}\,) + i \epsilon} \, \frac{1}{q^2 - m^2_l + i \epsilon} \nonumber \\
&=& \int_{\mid {\vec q}\, \mid < q_{\rm max}} \, \frac{d^3 q}{(2 \pi)^3} \,
\frac{1}{2 \omega_l (\vec q\,)} \frac{M_l}{E_l (-\vec{q}\,)} \,
\frac{1}{\sqrt{s}- \omega_l (\vec{q}\,) - E_l (-\vec{q}\,) + i \epsilon} \, ,
\label{eq:gprop}
\end{eqnarray}
with $\omega_l$ and $E_l$ being the energy of the meson (baryon) in the intermediate state in the c.m. frame, respectively, and $q_{\rm max}=630$~MeV, which has been fixed in this scheme to reproduce the $\Lambda(1405)$ properties and several threshold branching ratios \cite{Oset:1997it}.

The main contribution to the $p$-wave comes from the $\Lambda$ and $\Sigma$ pole terms, which are obtained  from the $D$ and $F$ terms of the lowest-order meson-baryon chiral Lagrangian  \cite{Jido:2002zk}. The $\Sigma^*(1385)$, belonging to the baryon decuplet, is also accounted for explicitly in our approach. The coupling of the $\Sigma^*$ to the $\bar K N$ system and other channels was elaborated in \cite{Oset:2000eg} according to quark-model $SU(6)$ symmetry.
Due to its spin structure, the $p$-wave terms from the chiral Lagrangian contribute to both the $J=1/2$ and $J=3/2$ $p$-wave meson-baryon amplitudes, with $J$ the total angular momentum. In order to obtain the leading-order amplitudes for $J=1/2,3/2$ we proceed as follows. The general expression for the partial-wave expansion of the scattering amplitude of a spin zero meson and a spin $1/2$ baryon reads
\begin{eqnarray}
    f(\vecq\,^{\prime}, \vecq) & = &\sum_{L=0}^{\infty} \left\{ (L+1) 
    f_{L+} + L f_{L-} \right\} P_{L}(\cos \theta) \nonumber \\
    && -i \vecsig 
    \cdot ( \hat q^{\prime} \times \hat q) \sum_{L=0}^{\infty} \left\{ 
    f_{L+} - f_{L-} \right\} P^{\prime}_{L}(\cos \theta) \ ,
    \label{partwaveamp}
\end{eqnarray}
where $\vec{q}$~($\vec{q}\,'$) is the three-momentum of the incoming (outgoing) meson and $\theta=\angle(\vec{q},\vec{q}\,')$.
In the previous expression the separation into spin-non-flip and spin-flip parts is manifest and each partial-wave amplitude $f_{L\pm}$ corresponds to orbital angular momentum $L$ and total angular momentum $J=L\pm1/2$.
In particular, for $L=1$ ($p$-wave interaction) one writes in a more usual notation
\begin{eqnarray}
V^p(\vecq\,^\prime, \vecq\,) = (2L +1) \lbrack f(\sqrt{s})\, \hat
 q^{\prime} \cdot \hat q - i g(\sqrt{s})\, (\hat q^{\prime} \times
 \hat q) \cdot \vecsig \rbrack  \label{pwamp} \ ,
\end{eqnarray}
where two amplitudes at tree level, $f_{-}^{\rm tree}$ ($L=1$, $J=1/2$) and
$f_{+}^{\rm tree}$ ($L=1$, $J=3/2$), can be defined as
\begin{eqnarray}
    f_{+}^{\rm tree} &=& f+g \label{fg} \\
    f_{-}^{\rm tree} &=& f-2g \nonumber \ ,
\end{eqnarray}
with
\begin{eqnarray}
    f_{ij}(\sqrt{s}) &=& {1 \over 3} \left\{ - C_{ij} {1 \over
4 f^2}\, a_i\,
    a_j \left({1 \over b_i} + {1 \over b_j} \right)
    + { D^{\Lambda}_i D^{\Lambda}_j \left(1+{q_i^0 \over M_i} \right)
    \left(1+{q_j^0 \over M_j} \right) \over \sqrt{s} - \tilde M_\Lambda}
    \right.  \nonumber \\
    && \left.  + { D^{\Sigma}_i D^{\Sigma}_j \left(1+{q_i^0 \over M_i}
    \right) \left(1+{q_j^0 \over M_j} \right) \over \sqrt{s} - \tilde
    M_\Sigma}
    + {2 \over 3} {D^{\Sigma^{*}}_i D^{\Sigma^{*}}_j \over
    \sqrt{s} - \tilde M_\Sigma^{*}} \right\} q_{i} q_{j}
    \label{f1}\\
    g_{ij}(\sqrt{s}) &=& {1 \over 3} \left\{  C_{ij} {1 \over 4
f^2}\, a_i\,
    a_j \left({1 \over b_i} + {1 \over b_j} \right)
    - { D^{\Lambda}_i D^{\Lambda}_j \left(1+{q_i^0 \over M_i} \right)
    \left(1+{q_j^0 \over M_j} \right) \over \sqrt{s} - \tilde M_\Lambda}
    \right.   \nonumber \\
    && \left.  - { D^{\Sigma}_i D^{\Sigma}_j \left(1+{q_i^0 \over M_i}
    \right) \left(1+{q_j^0 \over M_j} \right) \over \sqrt{s} - \tilde
    M_\Sigma} + {1 \over 3} {D^{\Sigma^{*}}_i D^{\Sigma^{*}}_j \over
    \sqrt{s} - \tilde M_\Sigma^{*}} \right\} q_{i} q_{j} \label{g1} \ ,
\end{eqnarray}
where $i,j$ are channel indices and $q_{i(j)}\equiv|\vec{q}_{i(j)}|$ here. The first term in both $f_{ij}$ and $g_{ij}$ comes from the small $p$-wave component in the meson-baryon amplitudes from the lowest order chiral Lagrangian in Eq.~(\ref{lowest}) \cite{Tolos:2006ny}, with
\begin{equation}
   a_i = \sqrt{E_i + M_i \over 2 M_i}\ , \hspace{0.7cm} b_i = E_i +
   M_i\ , \hspace{0.7cm} E_i = \sqrt{M_i^{\, 2} + \vec q_i\,^{ 2}} \  ,
\end{equation}
given in the c.m. frame. Moreover, the couplings of the hyperons excited in the $p$-wave amplitude to a given meson-baryon pair in channel $i$, $D^Y_i$, read
\begin{eqnarray}
   D^\Lambda_i &=& c_i^{D,\Lambda} \sqrt{20 \over 3} {D \over 2 f} -
   c_i^{F,\Lambda} \sqrt{12} { F \over 2 f} \nonumber \ , \\
   D^\Sigma_i &=& c_i^{D,\Sigma} \sqrt{20 \over 3} {D \over 2 f} -
   c_i^{F,\Sigma} \sqrt{12} { F \over 2 f} \ , \\
   D^{\Sigma^*}_i &=& c_i^{S,\Sigma^*} {12 \over 5} {D + F\over 2 f}
   \nonumber \ .
\end{eqnarray}
The constants $c^D$, $c^F$, $c^S$ are given by the pertinent $SU(3)$ Clebsch-Gordan coefficients and can be found in Table~I of Ref.~\cite{Jido:2002zk}, whereas the leading-order (vector and axial vector) meson-baryon chiral couplings $D$ and $F$ are chosen as $D=0.85$ and $F=0.52$. The masses $\tilde M_\Lambda$,
$\tilde M_\Sigma$, $\tilde M_{\Sigma^*}$ are bare masses of the
hyperons ($\tilde M_\Lambda$$=$1030 MeV, $\tilde M_\Sigma$$=$1120 MeV, $\tilde M_{\Sigma^*}$$=$1371 MeV),  which will turn into physical masses upon unitarization.

Unitarization proceeds in a similar way as described for the $s$-wave contribution. The on-shell factorization for $p$-waves in meson-baryon scattering is proven along the same lines as in meson-meson scattering \cite{Cabrera:2000dx}. Using Eq.~(\ref{eq:BS-matrix}), one obtains
\begin{eqnarray}
    f_{+} &=& [1-f_{+}^{\rm tree} G ]^{-1} f_{+}^{\rm tree}  \ ,
    \label{fs} \\
    f_{-} &=& [1-f_{-}^{\rm tree} G ]^{-1} f_{-}^{\rm tree} \ ,  \nonumber
\end{eqnarray}
where the $f^\pm$ amplitudes decouple within the Bethe-Salpeter equation and thus are unitarized independently.
The  $\Sigma^*$ pole  for $I=1$
is contained in the  $f_{+}$ amplitude while  the $f_{-}$ amplitude includes the $\Lambda$ and $\Sigma$ poles for
$I=0$ and $I=1$, respectively [cf.~Eqs.~(\ref{fg}-\ref{g1})].

Note that the amplitudes $f_{+}^{\rm tree}$, $f_{-}^{\rm  tree}$ in the diagonal meson-baryon channels contain the factor $\vecq\,^{2}$, with $\vecq$ being the on-shell c.m.  momentum of the meson in this channel. For transition matrix elements from channel $i$ to $j$ the corresponding factor is $q_{i}q_{j}$, where the energy and momentum of the meson in a certain channel are given by the expressions
\begin{equation}
    E_{i} = { s + m_{i}^{2} - M_{i}^{2} \over 2 \sqrt{s}} \ ;
  \hspace{0.5cm} q_{i} = \sqrt{E_{i}^{2} - m_{i}^{2}} \ ,
\end{equation}
which also provide the analytical extrapolation below the threshold of the channel, where $q_i$ becomes purely imaginary.

\subsection{$S=-1$ meson-baryon amplitudes in hot nuclear matter}

We next discuss how the model is modified to account for medium effects in hot and dense nuclear matter.
In order to obtain the effective $s$- and $p$-wave $\bar KN$ amplitudes (and related ones) in hot and
dense matter, the meson-baryon loop functions $G(\sqrt{s})$ have to be calculated at finite temperature and baryonic density, accounting for the in-medium propagators of the particles in the intermediate states.

One of the main sources of density and temperature dependence comes from the Pauli principle. This is
implemented by replacing the free nucleon propagator in the loop function by the
corresponding in-medium one. The other essential source  is related to the fact that all mesons and baryons in the intermediate loops interact with the nucleons of the Fermi sea and their properties are modified with respect to those in vacuum.

All these changes are straightforwardly implemented within the Imaginary Time Formalism (IFT), as extensively discussed  in Ref.~\cite{Tolos:2008di}. Applying the (finite-temperature) Feynman rules in this approach
the meson-baryon propagator reads \cite{Tolos:2008di}
\begin{eqnarray}
\label{G_ITF0}
{\cal G}_{MB}(W_m,\vec{P};T) &=& - T \int \frac{d^3q}{(2\pi)^3} \,
\sum_n 
{\cal D}_B(W_m-\omega_n,\vec{P}-\vec{q};T) \,
{\cal D}_M(\omega_n,\vec{q};T)
\ ,
\end{eqnarray}
where $T$ is the temperature, $\vec{P}$ is the external
total three-momentum, $\vec{q}$ the relative momentum and $W_m$ an external fermionic frequency,
${\rm i} W_m={\rm i} (2m+1)\pi T + \mu_B$, with $\mu_B$ being the baryonic chemical potential. The baryon and meson propagators within the Matsubara sum are given by
\begin{eqnarray}
\label{eq:DmesonDbaryon}
{\cal D}_B(w_n,\vec{p};T) &=& [{\rm i} w_n  - E_B(\vec{p},T)]^{-1} \ , \nonumber \\
{\cal D}_M(\omega_n,\vec{q};T) &=& [({\rm i} \omega_n)^2-\vec{q}\,^2 - m_M^2 -
\Pi_M(\omega_n,\vec{q};T)]^{-1} \ ,
\end{eqnarray}
with frequencies ${\rm i} w_n={\rm i} (2n+1)\pi T + \mu_B$ (fermionic) and ${\rm i} \omega_n = {\rm i} 2 \pi n T$ (bosonic). $E_B$ stands for the single-particle baryon energy and $\Pi_M$ denotes the pseudoscalar meson self-energy, which we discuss in more detail below.
The sum over the index $n$ is not straightforward because the meson self-energy depends on $n$ in a non-trivial way. This complication is circumvented by rewriting the meson propagator, $D_M$, in the spectral (Lehmann) representation, i.e.
\begin{eqnarray}
\label{Lehmann}
D_M(\omega_n,\vec{q};T) =
\int_0^{\infty} d\omega \,
\frac{S_M(\omega,\vec{q};T)}{{\rm i}\omega_n - \omega}
-
\int_0^{\infty} d\omega \,
\frac{S_{\bar M}(\omega,\vec{q};T)}{{\rm i}\omega_n + \omega}
\,\,\, ,
\end{eqnarray}
where $S_M$ and $S_{\bar M}$ stand for the spectral functions of the meson and its corresponding anti-particle. The relation between the meson spectral function and the propagator is evident by performing the analytical continuation from the Matsubara frequencies onto the real energy axis [$D_M(\omega,\vec{q};T)={\cal D}_M ({\rm i}\omega_n \to \omega+i\epsilon,\vec{q};T)$],
\begin{equation}
S_M(\omega,{\vec q}; T)= -\frac{1}{\pi} {\rm Im}\, D_M(\omega,{\vec q};T)
= -\frac{1}{\pi}\frac{{\rm Im}\, \Pi_M(\omega,\vec{q};T)}{\mid
\omega^2-\vec{q}\,^2-m_M^2- \Pi_M(\omega,\vec{q};T) \mid^2 } \ .
\label{eq:spec}
\end{equation}
Replacing Eq.~(\ref{Lehmann}) in Eq.~(\ref{G_ITF0}) one has
\begin{eqnarray}
\label{G_ITF}
{\cal G}_{MB}(W_m,\vec{P};T) &=& - T \int \frac{d^3q}{(2\pi)^3} \,
\sum_n \frac{1}{{\rm i} W_m - {\rm i}\omega_n - E_B(\vec{P}-\vec{q},T)} \,
\nonumber \\
&\times&
\int_0^{\infty} d\omega \,
\left[ \frac{S_M(\omega,\vec{q};T)}{{\rm i}\omega_n - \omega}
- \frac{S_{\bar M}(\omega,\vec{q};T)}{{\rm i}\omega_n + \omega} \right]
\,\,\, .
\end{eqnarray}
In this form the analytical structure of the meson-baryon loop is explicit and the Matsubara sums can be solved by using standard complex analysis techniques, leading to
\begin{eqnarray}
\label{G_ITF:Matsu-summed}
{\cal G}_{MB}(W_m,\vec{P};T) &=&
\int \frac{d^3q}{(2\pi)^3} \,
\int_0^{\infty} d\omega \,
\left[ S_M(\omega,\vec{q};T) \,
\frac{1-n_B(\vec{P}-\vec{q},T)+f(\omega,T)}
{{\rm i} W_m - \omega - E_B(\vec{P}-\vec{q},T)} \right.
\nonumber \\
&+&
\left.
S_{\bar M}(\omega,\vec{q};T) \,
\frac{n_B(\vec{P}-\vec{q},T)+f(\omega,T)}
{{\rm i} W_m + \omega - E_B(\vec{P}-\vec{q},T)} \, \right]
\,\,\, .
\end{eqnarray}
The properties of baryons in hot dense matter are implemented in the meson-baryon propagator in a two-fold manner. On the one hand, Pauli blocking is taken into account by considering the term 1-$n_B(\vec{P}-\vec{q},T)$, where $n_B(\vec{p},T)=[1+\exp(E_B(\vec{p},T)-\mu_B)/T)]^{-1}$ is the baryon Fermi-Dirac distribution. The single-particle baryon energy $E_B$ contains the medium binding effects obtained within a temperature dependent Walecka-type $\sigma -\omega$ model (see Ref.~\cite{KAP-GALE}). These binding effects are thus also present in the energy denominators.

The medium modifications on mesons, such as pions and antikaons, are incorporated in the meson-baryon loop by means of the inclusion of the meson Bose-Einstein distribution at finite temperature,  $f(\omega,T) = [\exp (\omega / T) - 1]^{-1}$, as well as  the meson  and its corresponding anti-particle spectral functions, $S_M(\omega,\vec{q};T)$ and $S_{\bar M}(\omega,\vec{q};T)$, defined above.
We consider in this work the dressing of pion and kaon propagators as they participate in the most relevant channels driving the meson-baryon interaction and the dynamical generation of the $\Lambda(1405)$.
For pions, we refer to Ref.~\cite{Tolos:2008di} for a detailed calculation of the pion self-energy at finite temperature within the ITF in the $ph-\Delta h$ model, including relativistic kinematics as well as full analyticity and crossing properties. For antikaons, the self-energy receives contributions of comparable size from both $s$- and $p$-wave interactions with the baryons in the medium. We refer the reader to the end of this section for details about its calculation.

The expression of Eq.~(\ref{G_ITF:Matsu-summed}) can be analytically continued onto the
real energy axis, $G_{MB}(P_0+{\rm i} \varepsilon \, ,\vec{P}; T) = {\cal
G}_{MB}({\rm i} W_m \to P_0 + {\rm i} \varepsilon \, , \vec{P}; T )$, where $P=(P_0,\vec{P})$ is the total
two-particle momentum. Here we provide the detailed expressions for the in-medium loop functions on the real energy axis, where some simplifications are applicable for practical purposes.

For $\bar KN$ states one has
\begin{eqnarray}
\label{eq:gmed}
{G}_{\bar KN}(P_0+{\rm i} \varepsilon,\vec{P};T)
&=&\int \frac{d^3 q}{(2 \pi)^3}
\frac{M_N}{E_N (\vec{P}-\vec{q},T)} \nonumber \\
&\times  &\left[ \int_0^\infty d\omega
 S_{\bar K}(\omega,{\vec q};T)
\frac{1-n_N(\vec{P}-\vec{q},T)}{P_0 + {\rm i} \varepsilon - \omega
- E_N
(\vec{P}-\vec{q},T) } \right. \nonumber \\
&+& \left. \int_0^\infty d\omega
 S_{K}(\omega,{\vec q};T)
\frac{n_N(\vec{P}-\vec{q},T)} {P_0 +{\rm i} \varepsilon + \omega -
E_N(\vec{P}-\vec{q},T)} \right] \ ,
\end{eqnarray}
with ${\vec q}$ being the meson three-momentum\footnote{We note the additional factor $M_B/E_B$ with respect to Eq.~(\ref{G_ITF:Matsu-summed}) in order to keep consistency with the normalization of the baryon propagator in free space.}. The second term in the $\bar KN$ loop function typically provides a small, real contribution for the studied energy range in $P_0$.
Here one can
replace $S_{K}(\omega, \vec q;T )$ by a free-space delta function, which simplifies numerical computations. The latter is a sensible approximation since the $K$ spectral function in the medium still peaks at the quasi-particle energy and the latter does not differ much from the energy in vacuum \cite{Tolos:2008di}. In addition, one finds that the kaon distribution function can be safely neglected at the temperatures of interest (we expect Bose enhancement to be relevant only for pions at $T = 0$ -- $150$~MeV \cite{Tolos:2008di}).

In the case of $\pi \Lambda$ or $\pi \Sigma$ states one gets
\begin{eqnarray}
\label{eq:gmed_piY}
{G}_{\pi Y}(P_0+{\rm i} \varepsilon,\vec{P}; T)
&= & \int \frac{d^3 q}{(2 \pi)^3} \frac{M_{Y}}{E_{Y}
(\vec{P}-\vec{q},T)} \nonumber \\
& \times &
\int_0^\infty d\omega
 S_\pi(\omega,{\vec q},T)
\left[
\frac{1+f(\omega,T)}
{P_0 + {\rm i} \varepsilon - \omega - E_{Y}
(\vec{P}-\vec{q},T) }   \right.
\nonumber \\
& + &
\left.
\frac{f(\omega,T)}
{P_0 + {\rm i} \varepsilon + \omega - E_{Y}
(\vec{P}-\vec{q},T) } \right] \ .
\end{eqnarray}
The $\pi Y$ loop function incorporates the $1+f(\omega ,T)$
enhancement factor which accounts for the  contribution from thermal pions at
finite temperature.
In this case, we have neglected the fermion distribution for the participating
hyperons, which is a reasonable approximation for the range of temperatures and
baryonic chemical potentials.

Finally, for $\eta \Lambda$, $\eta \Sigma$ and $K \Xi$ intermediate states,
we simply consider
the meson propagator in vacuum and include only the effective baryon
energies modified by the mean-field binding potential for $\Lambda$ and $\Sigma$ hyperons, i.e.
\begin{eqnarray}
G_i(P_0+{\rm i} \varepsilon,\vec{P};T)= \int \frac{d^3 q}{(2 \pi)^3} \,
\frac{1}{2 \omega_i (\vec q\,)} \frac{M_i}{E_i (\vec{P}-\vec{q},T)} \,
\frac{1}{P_0 +
{\rm i} \varepsilon - \omega_i (\vec{q}\,) - E_i (\vec{P}-\vec{q},T) } \, .
\label{eq:gmed-etaY-KXi}
\end{eqnarray}
This approximation is justified as the latter channels are less relevant in the unitarization procedure  \cite{Oset:1997it}.

In order to compute the in-medium $s$- and $p$-wave amplitudes of $\bar K N$ at finite temperature, one needs to solve Eq.~(\ref{eq:BS-matrix}) in matter. The on-shell factorization of the amplitudes in the Bethe-Salpeter equation can be maintained in the case of the in-medium calculation for $s$-wave scattering \cite{Tolos:2006ny}.  The amplitudes in the $p$-wave, however, require a slightly different treatment since the on-shell factorization is not exactly reproduced in the medium due to remaining tadpole contributions \cite{Tolos:2006ny}. As it was shown in Ref.~\cite{Tolos:2006ny}, the formal algebraic solution of the Bethe-Salpeter equation with on-shell amplitudes can be kept for the $p$-waves with a simple modification of the meson-baryon loop function, modulo some small tadpole corrections. Summarizing the results in Ref.~\cite{Tolos:2006ny}: if we denote by $G_i^L(P^0,\vec{P};T)$ the in-medium meson-baryon propagator for $s$- ($L=0$) and $p$-wave ($L=1$) scattering (and $i$ labels a specific 
$MB$ channel), one has:
\begin{eqnarray}
\label{eq:Gsummary}
G_i^{(s)}(P^0,\vec{P};T) &=& G_{i}(P_0+{\rm i} \varepsilon \, ,\vec{P}; T) \ , \nonumber \\
G_i^{(p)}(P^0,\vec{P};T) &=& G_i(s) + \frac{1}{\vec{q}\,^2_{\rm on}} [ \tilde{G}_{i}(P_0 +{\rm i} \varepsilon \,,\vec{P}; T) - \tilde{G}_i(s) ] \ ,
\end{eqnarray}
where the $\tilde{G}$ functions carry an extra $\vec{q}\,^2$ factor in the integrand, corresponding to the  off-shell $p$-wave vertex.

As discussed in \cite{Tolos:2006ny}, nuclear short-range correlations have to be taken into account when dealing with $p$-wave amplitudes in order to account for the fact that the nucleon-nucleon (hyperon-nucleon) interaction is not only driven by one-pion (one-kaon) exchange. These correlations arise when the $\pi$ ($\bar K$) in the meson-baryon loops are dressed in the medium and develop $NN^{-1}$ ($YN^{-1}$) excitations. The short-range part of the interaction is mimicked by phenomenological Landau-Migdal contact vertices ($NY$-$NY'$) and is technically implemented by replacing the propagator of the exchanged pion (kaon) in Eq.~(\ref{eq:DmesonDbaryon}) by a correlated interaction which performs the Dyson resummation of the irreducible meson selfenergy modified by successive iterations of the contact interaction [cf.~Eqs.~(30-35) in \cite{Tolos:2006ny} for detailed expressions].

Once the in-medium $\bar K N$ amplitudes at finite temperature are obtained, we can compute the $\bar K$ self-energy in either $s$- or $p$-wave by
integrating  the effective interaction $T_{\bar K N}$ over the nucleon Fermi distribution at a given
temperature, i.e.
\begin{eqnarray}
\Pi^L_{\bar K}(q_0,{\vec q};T)= 4 \int \frac{d^3p}{(2\pi)^3}\,
n_N(\vec{p},T) \,
\bar{T}^L_{\bar K N}(P_0,\vec{P};T) \ ,
\label{eq:selfd}
\end{eqnarray}
where $P_0=q_0+E_N(\vec{p},T)$ and $\vec{P}=\vec{q}+\vec{p}$ are
the total energy and momentum of the $\bar KN$ pair in the
nuclear medium rest frame, $q$ stands for the
momentum of the $\bar K$ meson also in this frame, and $\bar{T}$ indicates the spin and isospin averaged scattering amplitude for a given partial wave.
We also provide for convenience Eq.~(\ref{eq:selfd}) rewritten in the basis of physical states for antikaons,
\begin{eqnarray}
\Pi^L_{K^-}(q_0,{\vec q};T)= 2 \int \frac{d^3p}{(2\pi)^3}\,
\lbrack
n_p(\vec{p},T) \,
T^L_{K^-p}(P_0,\vec{P};T)
+
n_n(\vec{p},T) \,
T^L_{K^-n}(P_0,\vec{P};T)
\rbrack
\ ,
\label{eq:selfd-v2}
\end{eqnarray}
where the $L=1$ amplitude is defined as in Eq.~(\ref{pwamp}) and reads here $T^{L=1}=3\, \lbrack f_- + 2f_+ \rbrack$, with $f_\pm$ given in Eqs.~(\ref{fg}-\ref{g1},\ref{fs}). A similar expression is obtained for $\bar K^0$ and we recall that $\Pi_{K^-}=\Pi_{\bar K^0}\equiv\Pi_{\bar K}$ in symmetric nuclear matter.
The antikaon self-energy must be determined self-consistently since it is
obtained from the in-medium amplitude, $ T^L_{\bar K N}$, which
requires the evaluation of the $\bar KN$ loop function,
$G^L_{\bar KN}$, and the latter itself is a function of
$\Pi_{\bar K}(q_0, \vec q; T)$ through the antikaon spectral
function, cf.~Eqs.~(\ref{eq:spec}), (\ref{eq:gmed}).
Note that Eq.~(\ref{eq:selfd}) corresponds to a na\"{\i}ve generalization of the zero-temperature result, as discussed in Ref.~\cite{Tolos:2008di}. For completeness we provide a detailed derivation of the finite temperature antikaon self-energy in terms of the $\bar K N$ $T$-matrix in the appendix.

\section{Results for $S=-1$ meson-baryon amplitudes and hyperon single-particle properties in matter}
\label{sec:amplitudes}		

We discuss in the following our results for the scattering amplitudes in the isospin channels $I=0,1$ and  $s$- and $p$- waves at finite nuclear density and temperature. This information is accessible due to the extension of our model to account for unitarized amplitudes in both $s$- and $p$-waves and different isospin and $J^P$ channels. The final goal is to study the excited hyperon resonances and assess how the nuclear environment influences their properties.

\begin{figure}[t]
\begin{center}
\includegraphics[width=0.49\textwidth,height=0.5\textwidth]{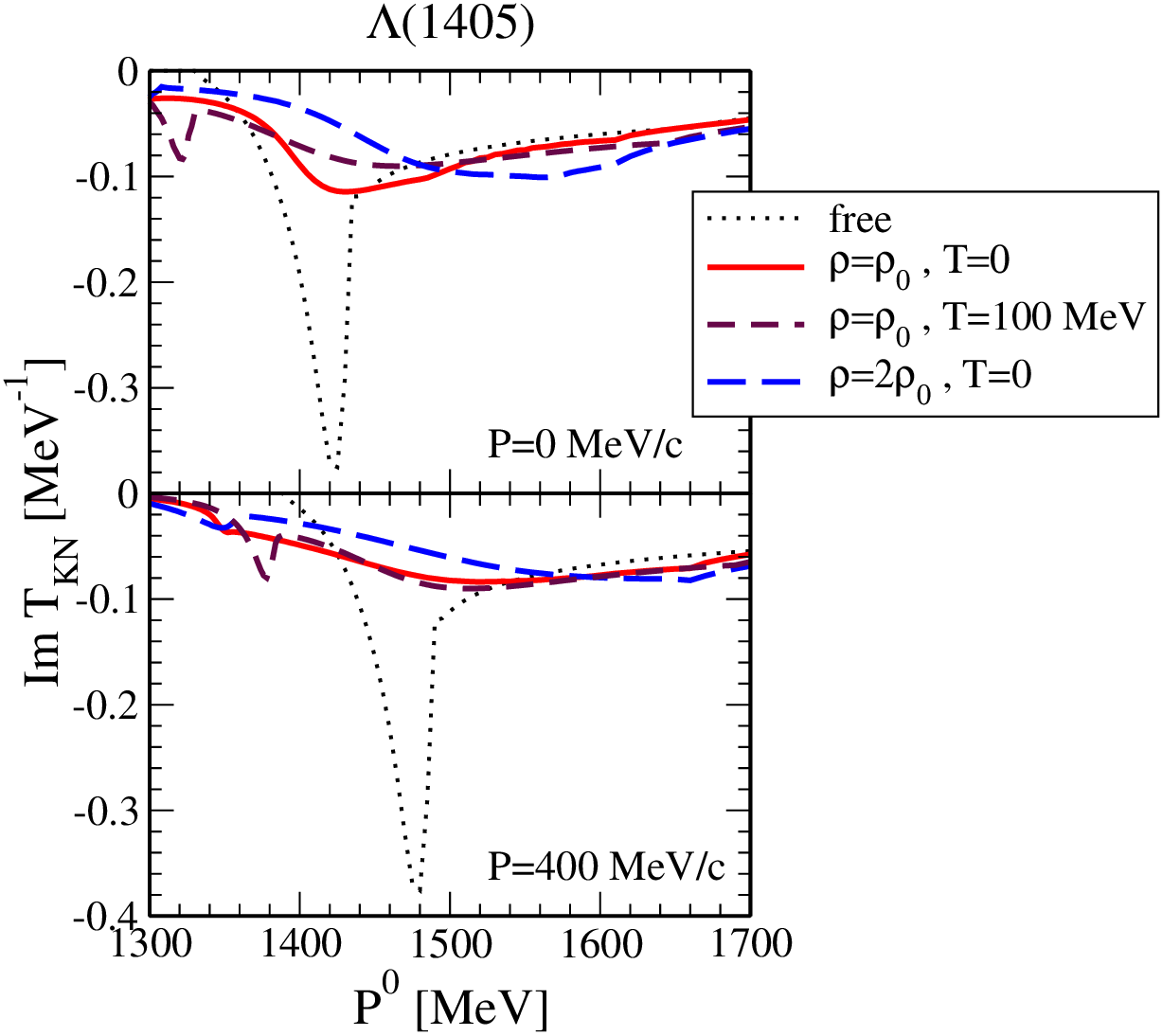}
\includegraphics[width=0.49\textwidth,height=0.5\textwidth]{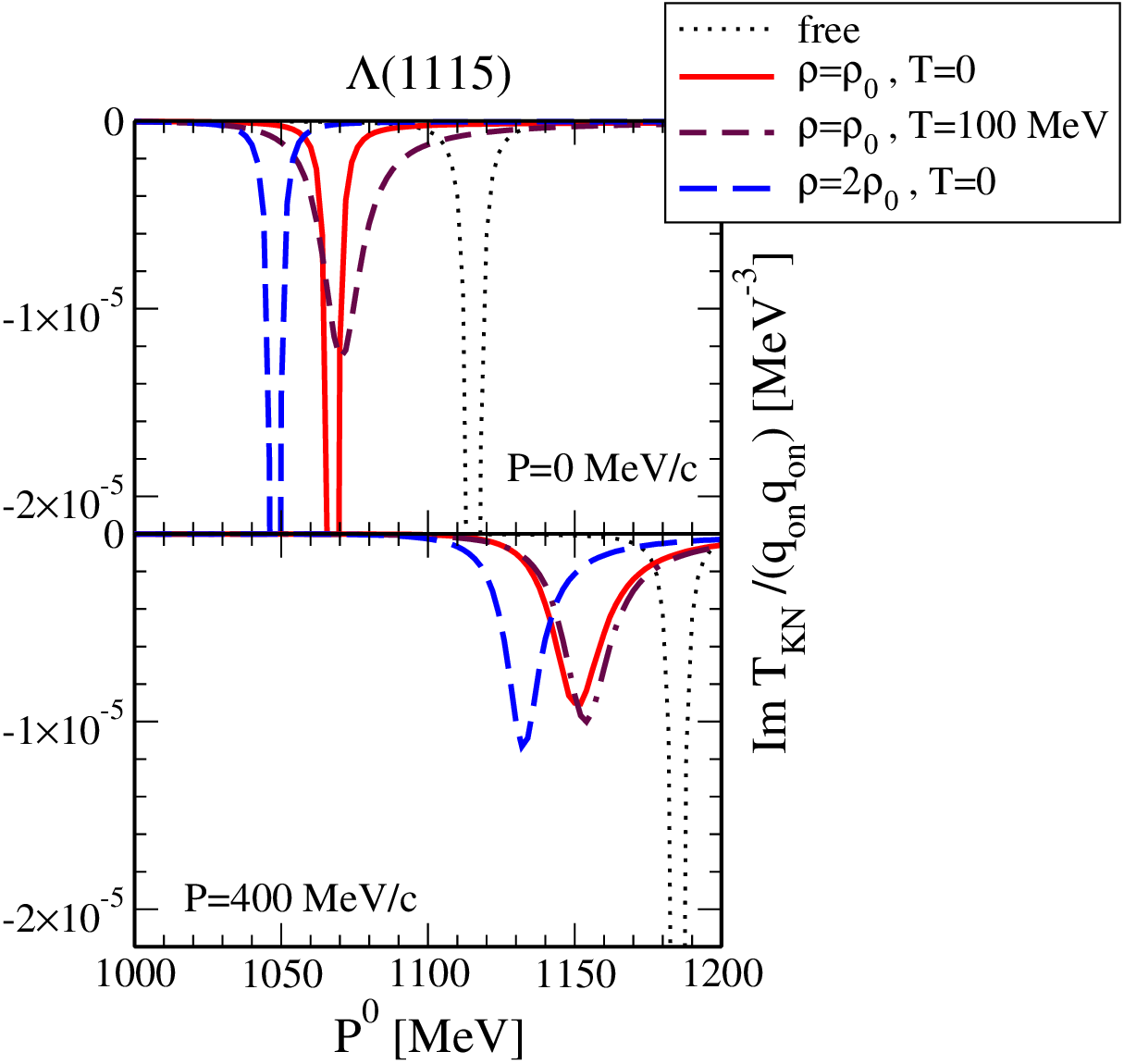}
\caption{Imaginary part of the $\bar K N$ scattering amplitude in vacuum and in the medium for specific resonant channels. Left: $I=0$, $L=0$ and $J=1/2$ amplitude ($\Lambda (1405)$ channel). Right: $I=0$, $L=1$ and $J=1/2$ amplitude ($\Lambda (1115)$ channel).}
\label{fig:iso-amp-lambdas}
\end{center}
\end{figure}

In Fig.~\ref{fig:iso-amp-lambdas} we depict the imaginary part of the $\bar K N \to \bar K N$ scattering amplitude in the isoscalar channel with $J^P=1/2^+$, for $L=0$ ($\Lambda(1405)$ channel, left panel) and $L=1$ ($\Lambda(1115)$ channel, right panel). We show two different values of the meson-baryon total momentum (upper and lower panels).\footnote{We note here that our results are available in a full $(P^0,P)$ grid.} We reproduce our previous results for the $\Lambda(1405)$ at nuclear saturation density and zero temperature \cite{Tolos:2006ny,Tolos:2008di}. This resonance strongly dilutes in the nuclear medium mostly due to the pion-related decay channels such as $\Lambda(1405)\to \Lambda NN^{-1},\Sigma NN^{-1}$ and similarly with $\Delta N^{-1}$ components, whereas the peak of the distribution (from here on, the quasi-particle energy) remains  slightly above its vacuum position for normal nuclear matter density, $\rho_0$. At $\rho=2\rho_0$ the distribution is substantially broader and appreciably 
shifted to higher energies. The effect of the temperature is two-fold: first, it further broadens the resonance as a result of the smearing of the Fermi surface, which increases the available phase space for in-medium decays. Second, the attractive baryonic potentials entering the quasi-particle energies of nucleons and hyperons in the meson-baryon loops become shallower with increasing the temperature, implying that all meson-baryon thresholds are shifted to higher energies with respect to the $T=0$ case. This can be easily appreciated in Fig.~\ref{fig:iso-amp-lambdas} (left), where at $T=100$~MeV the $\Lambda(1405)$ is dynamically generated at a slightly higher $\sqrt{s}$ and the kink corresponding to the opening of the in-medium $\pi\Sigma$ channel, which remains below the range in the plot at $T=0$, is visible at the low-energy tail of the resonance at finite temperature.

The $\Lambda(1115)$ exhibits attraction in the nuclear medium, which in our approach amounts to about -$48$~MeV at normal nuclear matter density, and is essentially dominated by the pion mediated $\Lambda N \to \Sigma N$ transition incorporated in our approach by the dressing of pions and the implementation of short-range correlations (the lack of the latter leads to unphysically larger attractive shifts by roughly a factor $\sim$2). We note that the apparent width of the resonance at $P=0$ and zero temperature is simply a numerical artifact in order to solve the matrix inversion problem, whereas at finite total momentum the resonance acquires a physical finite width from intermediate $\Lambda NN^{-1}$ excited states. At finite temperature, however, the broadened Fermi distribution of nucleons allows to accommodate such excitations even at $P=0$, and the $\Lambda(1115)$ develops a finite decay width as can be seen in the right panel of Fig.~\ref{fig:iso-amp-lambdas} for the $T=100$~MeV case, whereas the 
attraction on the quasi-particle energy is slightly reduced.
The attractive shift at $T=0$ found here for the $\Lambda(1115)$ overestimates previous determinations within the same model in \cite{Tolos:2006ny} and meson exchange models \cite{Reuber:1993ip,Stoks:1999bz,Rijken:1998yy,Vidana:2001rm,Haidenbauer:2005zh}, which estimate an attraction for the $\Lambda$ in nuclear matter of about -$30$~MeV at $\rho=\rho_0$, as required by hypernuclear spectroscopy \cite{hyper}. 
Our larger shift is partly due to the input baryonic mean-field potential for the hyperons ($\Lambda$ and $\Sigma$), which are estimated from those of the nucleon within a $\sigma$-$\omega$ model at finite density and temperature by means of simple quark-model counting rules. The model leads to an attractive binding for both hyperons of approximately -$50$~MeV at $\rho=\rho_0$. We have used this model as it incorporates the temperature dependence of the baryonic potentials and also in order to compare the present results to our previous calculation  \cite{Tolos:2008di}, where $p$-wave unitarization at finite density and temperature was missing.
The hyperon binding potential can be readily improved by modifying the scalar ($\sigma YY$) and vector ($\omega YY$) couplings ($g_{\sigma/\omega YY} = \alpha\,  g_{\sigma/\omega NN}$ with $\alpha$=$2/3$ within the strict quark counting scheme) so as to satisfy the phenomenological requirement $U_\Lambda(\rho_0)\simeq$-$30$~MeV.
We find, however, that such modifications barely affect the $\Lambda$ and $\Sigma$ mass shifts obtained from the $p$-wave amplitudes, indicating that 
the effect of the input baryonic potentials saturates to some extent within our self-consistent calculation.
On top of this, the impact of these variations is marginal on the position and  shape of the $\Lambda(1405)$ resonance. Therefore, although the binding potentials certainly influence the eventual nuclear potential of the hyperons in the $p$-waves, one can say that it is not the leading effect in our calculation.
As discussed before, the former test corroborates that the attractive potentials that the $\Lambda$ and $\Sigma$ develop at finite density are mostly due to the pion-mediated coupled channels, when the pion is also dressed in the medium and short-range correlations within vertices related to the $NN$ and $NY$ interaction are simultaneously implemented \cite{Tolos:2006ny}. The strength of these mechanisms depend on a reduced set of parameters, namely the baryonic form factor of the pion (with scale parameter $\Lambda_{\pi}$), accounting for the finite size of $\pi NN$,~$\pi N\Delta$ vertices; and the Landau-Migdal parameter, $g'$, controlling the size of short-range correlations.
We have checked that varying these parameters within realistic ranges ($\Lambda_{\pi}\simeq 0.8$-$1$~GeV, $g'\simeq 0.6$-$0.8$) one can accommodate the value of $U_\Lambda(\rho_0)\simeq$-$30$~MeV. This can be achieved by using a softer hadronic pion form factor, with $\Lambda_{\pi}\simeq 0.8$~GeV, and $g'\simeq0.6$. For this set of parameters the nuclear potential for the $\Sigma$ is reduced to approximately -$25$~MeV at $\rho_0$ (note that the $\Sigma$ is even more sensitive to the pion properties due to the in-medium open channels $\Sigma N \to \Sigma N, \Lambda N$).
The need for a softer pion form factor in our calculation, as compared to previous studies within similar models in cold nuclear matter, can be justified as follows: within heavy-ion studies, non-relativistic approximations typically performed to simplify the calculation of $NN^{-1}$, $\Delta N^{-1}$, $YN^{-1}$ excitation functions (Lindhard-Migdal functions) are not suitable since the meson-baryon pair scans a larger set of states in momentum space. The use of fully relativistic kinematics in the baryon propagators \cite{Tolos:2008di} results in meson self-energies with slower high-energy and momentum behavior, leading to stronger effects from the in-medium pion dressing. The use of a slightly softer hadronic form factor for the pion selfenergy is enough to compensate this extra strength from pion-related coupled channels.

\begin{figure}[t]
\begin{center}
\includegraphics[width=0.49\textwidth,height=0.5\textwidth]{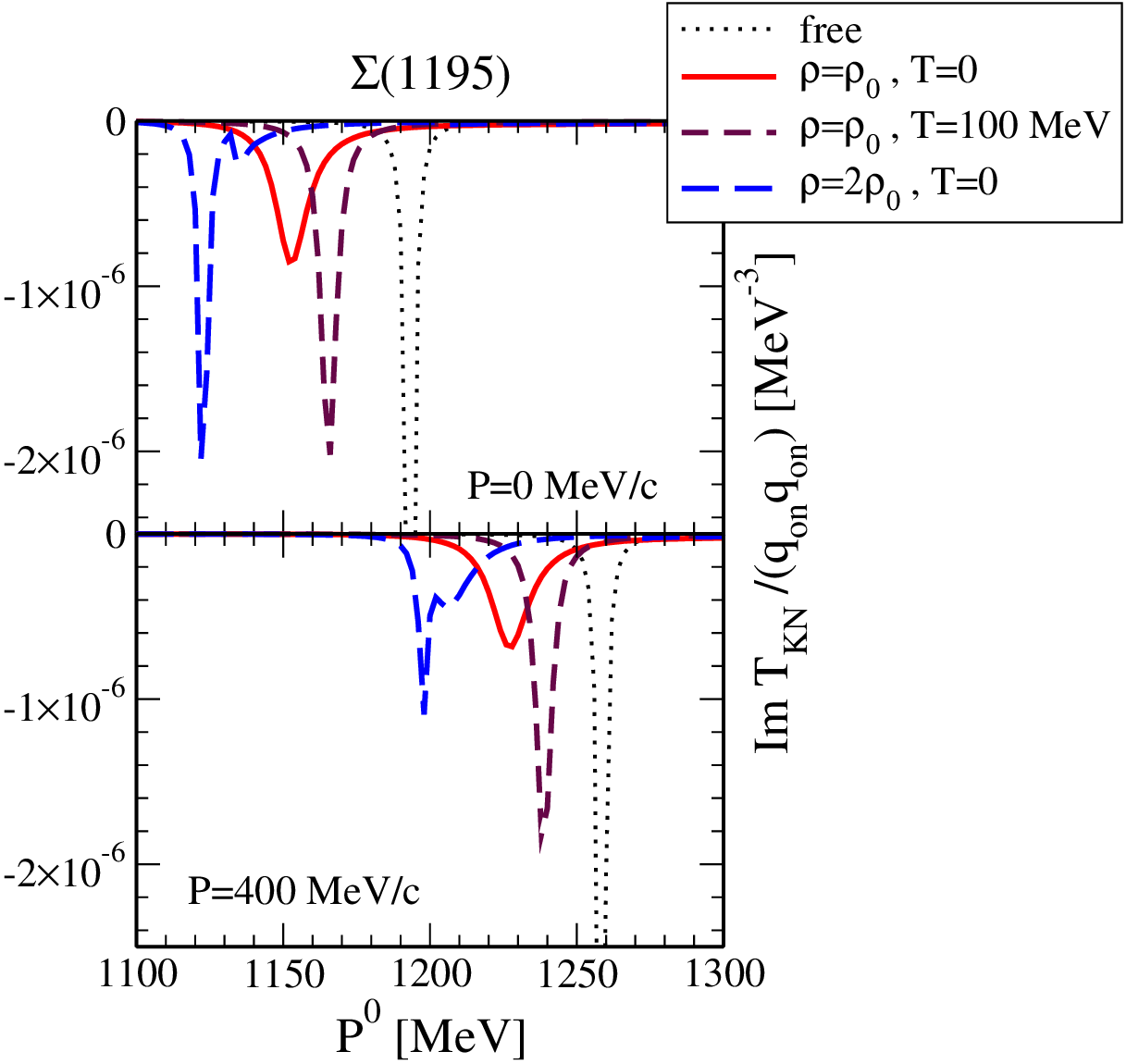}
\includegraphics[width=0.49\textwidth,height=0.5\textwidth]{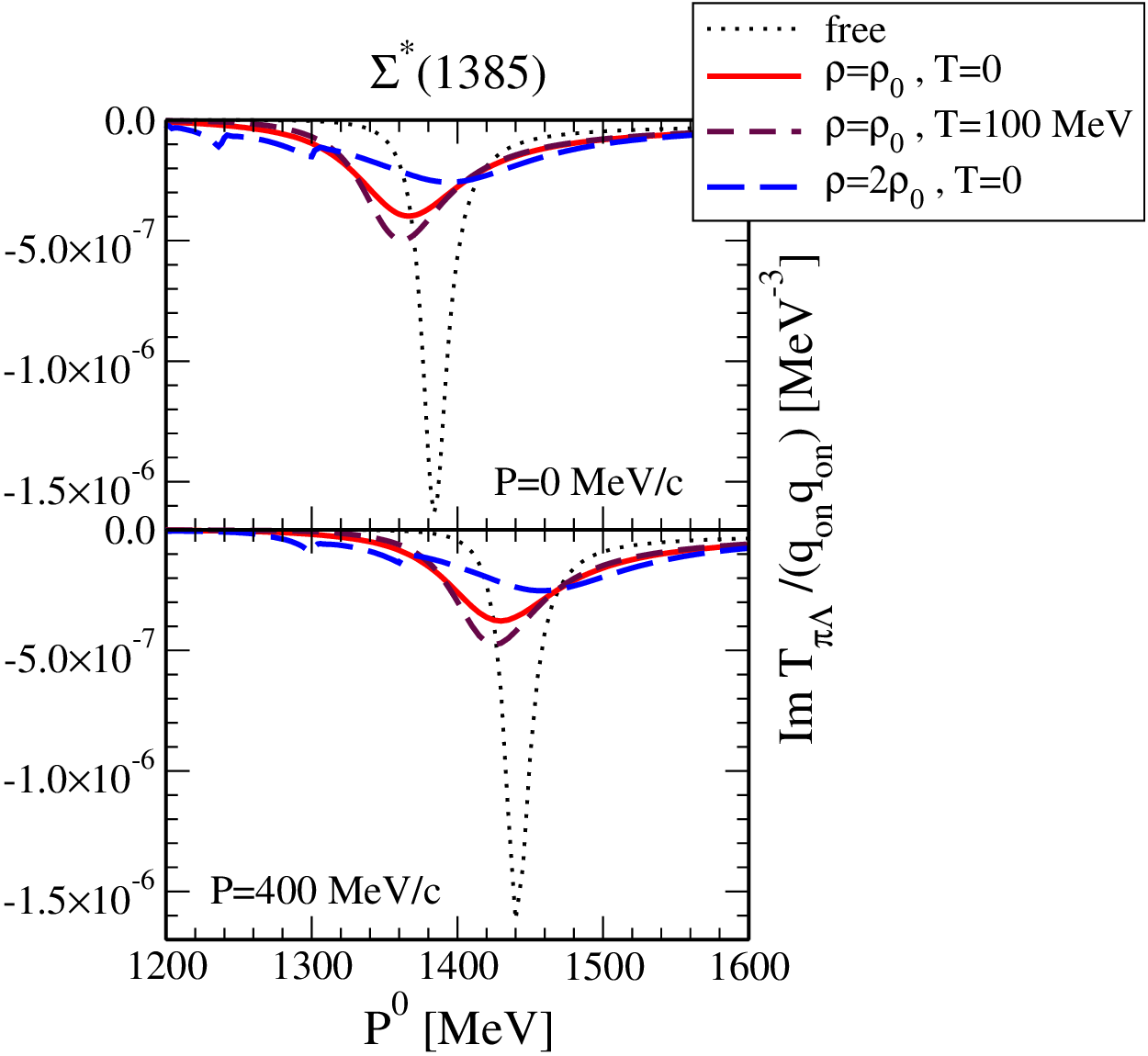}
\caption{
Same as in Fig.~\ref{fig:iso-amp-lambdas} for the isovector hyperon channels.
Left: $I=1$, $L=1$ and $J=1/2$ $\bar K N$ amplitude ($\Sigma (1195)$ channel). Right: $I=1$, $L=1$ and $J=3/2$ $\pi\Lambda$ amplitude ($\Sigma^* (1385)$ channel).}
\label{fig:iso-amp-sigmas}
\end{center}
\end{figure}

Our results for the isovector hyperons are shown in Fig.~\ref{fig:iso-amp-sigmas}. The left panel corresponds to the $\bar K N \to \bar K N$ $p$-wave amplitude for $J^P=1/2^+$, where the $\Sigma(1195)$ is excited, whereas the right panel shows the $J^P=3/2^+$ component of the $p$-wave $\pi \Lambda\to\pi\Lambda$ amplitude, dominated by the $\Sigma^*(1385)$. The $\Sigma$ acquires an attractive shift of about -$40$~MeV at normal matter density, about $5$~MeV larger than in \cite{Tolos:2006ny} and again mostly due to the pion-mediated $YN$ interaction.
Its decay width at $P=0$ originates from the $\Sigma N\to\Lambda N$ transition, readily incorporated in the model through the dressing of pions and kaons. At $T=100$~MeV, the $\Sigma$ attraction is reduced by about $1/3$ of the value at zero temperature. The $\Sigma$ also becomes narrower, which seems counterintuitive given the expected enhancement of phase space from a broader nucleon distribution at finite temperature. However, one should keep in mind that the baryons in all intermediate states also become heavier with increasing temperature and that the mass difference between the $\Sigma$ and $\Lambda$ hyperons becomes smaller with temperature. Given the relatively small energies available for $\Sigma\to\Lambda N N^{-1}$ decays the latter effect dominates and the $\Sigma$ width is reduced. At $\rho=2\rho_0$, the $\Sigma$ profile displays a kink at $\sim 1140$~MeV and peaks below this energy. This is due to the large attraction which shifts the $\Sigma$ state below the in-medium $\pi \Sigma$  threshold, 
and consequently the hyperon in-medium width is reduced. This effect would have been smeared out if we had performed a self-consistent calculation  for the hyperon single-particle potential in dense matter.
Our present results regarding the $\Sigma$ self-energy in the medium are comparable to former determinations \cite{Batty:1978sb,pedro} in cold nuclear matter. Other approaches based on phenomenological potentials constrained by $\Sigma$-atom data conclude that the $\Sigma$ experiences repulsion at short distances, while the potential turns attractive at large distances \cite{Batty:1994yx,Mares:1995bm}.
It is worth mentioning the model calculation of \cite{Kaiser:2005tu}, based on the meson-baryon chiral Lagrangian and accounting for long-range dynamics (pion and kaon exchange mechanisms), which finds a net repulsive potential of about $60$~MeV at nuclear matter density.
The theoretical status of the $\Sigma$ potential seems to be far from being settled, whereas the only experimental evidence is that $\Sigma$-atoms require an
attraction at the relatively large distances that are probed in these experiments.
The study of inclusive spectra in $(\pi^-, K^+)$ $\Sigma$-production reactions provides complementary information. In \cite{Noumi:2001tx,Saha:2004ha,Kohno:2004pb} these spectra have been analyzed within the
distorted wave approximation for pions and kaons (see also \cite{Morimatsu:1994sx}, where the equivalent Green's function method is used) with the conclusion of a repulsive
$\Sigma$-nucleus potential at central densities.
These results, however, should be pondered with care since the method employed may not be appropriate for inclusive reactions (where one sums over all possible nuclear final states), as the distorted wave approximation removes $K$ and $\pi$ quasi-elastic and absorption events from the flux whereas the resulting final state still contains the particles of interest. Presumably, this method forces a repulsive $\Sigma$-nucleus potential in order to 
prevent the $\Sigma$ hyperon from being too close to the nucleus (and scan larger densities), as the distorted pion and kaon waves would then remove too many events from the flux.
Summarizing, in our understanding the experimental situation concerning the $\Sigma$ potential is also unresolved and, again, the only robust information that can be presently extracted is that the $\Sigma$ potential is attractive at the small densities probed in atom production.

The $\Sigma^*(1385)$ has a finite decay width in vacuum from $\pi \Lambda$ and $\pi \Sigma$ decays, which is correctly accounted for in our model. At finite nuclear density, the opening of additional decay channels related to the pion and kaon dressing (thus $\Sigma^* \to \Lambda N N^{-1}, \Sigma N N^{-1}$) and the $p$-wave character of the interaction enhances considerably the $\Sigma^*$ width, which evolves from $35$~MeV in vacuum to close to $100$~MeV at $\rho=\rho_0$. This value has to be compared to $80$~MeV as obtained in \cite{Tolos:2006ny,Kaskulov:2005uw}; the larger value in our case is related to the more attractive potentials acting on the $\Lambda$ and $\Sigma$ hyperons as well as the increase of the $\Sigma^*$ mass with density. Indeed, the attractive $\Sigma^*$ mass shift of $10$~MeV  at $\rho=\rho_0$  turns into repulsion at larger densities. At two times normal matter density the resonance is so broad that it starts to be meaningless to define a quasi-particle energy or a mass shift. The 
effect of the temperature in this case is moderate due to the important phase space already available  at zero temperature. Due to the baryons picking up larger quasi-particle energies and the slight reduction of the $\Sigma^*$ mass, the $\Sigma^*$ width is actually mildly reduced at $\rho=\rho_0$ and $T=100$~MeV as compared to the zero temperature case.

\begin{figure}[t]
\begin{center}
\includegraphics[height=6cm]{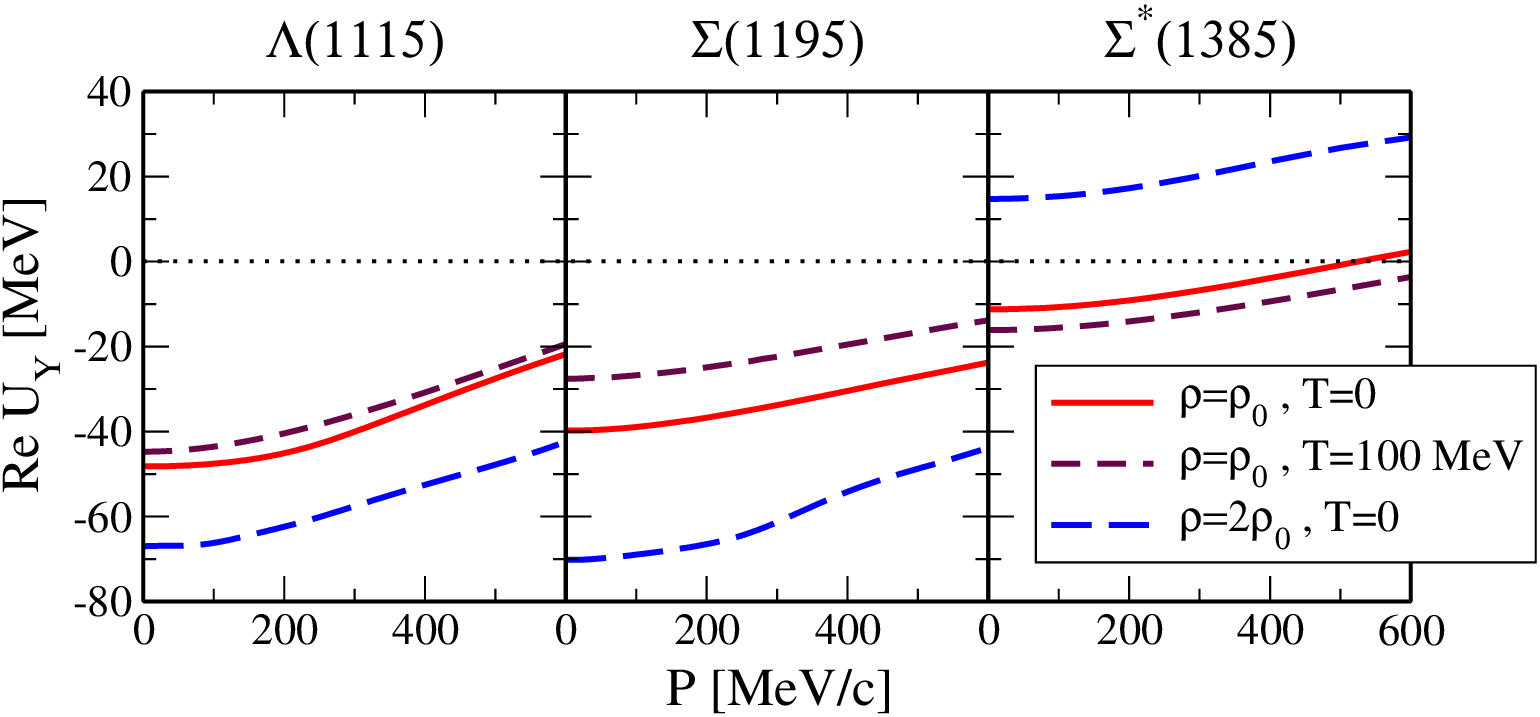}\\
\hspace{-0.9cm}
\includegraphics[height=6cm]{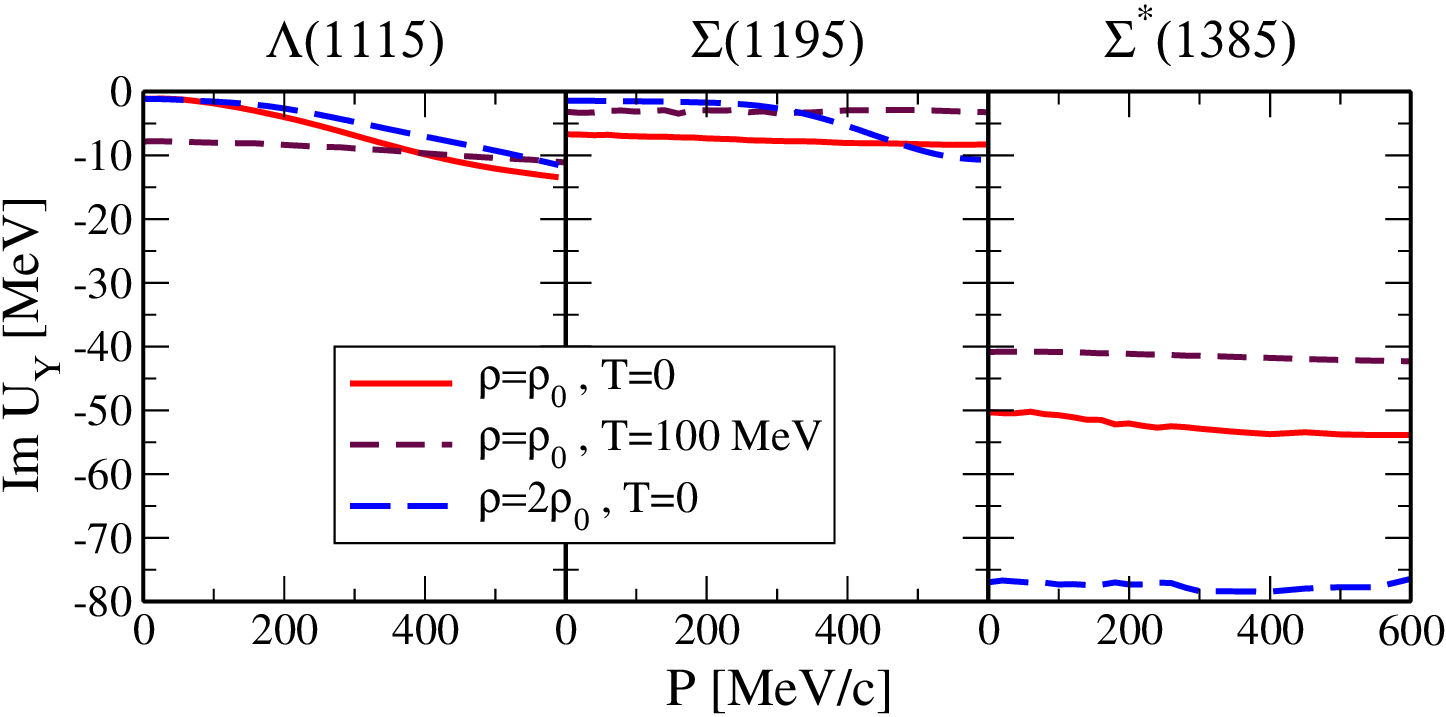}
\caption{Momentum dependence of the nuclear optical potential for the  $\Lambda(1115)$, $\Sigma(1195)$ and $\Sigma^*(1385)$ hyperons  at finite nuclear density and temperature. Upper panel: real part. Lower panel: imaginary part.}
\label{fig:UY}
\end{center}
\end{figure}

It is pertinent to make a comparison of our results with similar approaches, particularly that in Ref.~\cite{Lutz:2007bh} by Lutz~{\it et al.}, where a self-consistent and covariant many-body approach based on the chiral $SU(3)$ Lagrangian is employed to study antikaon dynamics in dense nuclear matter at zero temperature.
The most relevant differences can be summarized as follows: a) the angular integration in the meson-baryon loop function [cf.~Eq.~(\ref{G_ITF:Matsu-summed})] is approximated in our calculation by an average over the Fermi distribution, whereas in \cite{Lutz:2007bh} it is also evaluated explicitly. b) We incorporate density and temperature dependent scalar and vector mean-field potentials for the nucleon and the ground-state hyperons, whereas in \cite{Lutz:2007bh} this is only implemented for the nucleon. c) The interaction of the hyperons with the $\bar K N$ system in the $p$-wave amplitudes is modified by short-range correlations in our approach, in consistency with the phenomenology of nucleon-nucleon and hyperon-nucleon interactions, which requires a treatment of short distances beyond the one-pion and one-kaon exchange mechanisms.
The effect of the angular average has been analyzed in \cite{Lutz:2007bh} and the authors conclude that the impact of this approximation is marginal for the antikaon spectral function, as previously stated in \cite{Ramos:1999ku}, whereas it becomes more important for the contribution of $d$-wave interactions, driven by the excitation of the $\Lambda(1520)$ (not accounted for in our model).
Still, some differences are found when comparing the scattering amplitudes in the $p$-wave and the in-medium excitation energy of the $\Lambda$, $\Sigma$ and $\Sigma^*$, which, disregarding the effect of the angular average, can only be adscribed to the different strength of the $p$-wave interaction in both approaches (short-range phenomena are not implemented in the calculation of Lutz~{\it et al.})
Overall, the size of the attraction experienced by the $\Lambda$ and $\Sigma$ in \cite{Lutz:2007bh} is larger than in our case (by roughly factor $1.5$-$2$), a feature that we can also reproduce if short-range interactions are switched off. The discrepancy with the mass shift of the $\Sigma^*(1385)$ is even more dramatic, of the order of a factor $\sim$4.
Incidentally, a narrow soft mode associated to a highly collective $\Lambda N^{-1}$ excitation is observed by the authors of Ref.~\cite{Lutz:2007bh} in the low-energy tail of the $\bar K$ spectral function. Such a peaky structure is not present in our previous results \cite{Tolos:2006ny} and thus one can infer that short-range correlations in the $\bar K N$ interaction are taming the strength of this low-energy mode.
The emergence of a low-energy tail in the $\bar K$ spectral function due to many-body correlations is an important phenomenon with direct connection with the possibility of formation of kaon condensates in dense matter (e.g.~in compact stars). Populated by $YN^{-1}$ excitations in the $p$-wave, and enhanced at finite temperature as discussed in \cite{Tolos:2008di}, such soft modes in the $\bar K$ spectral function are likely to increase the reactivity of the $\phi$ meson at FAIR energies by "stimulated" $\phi\to\bar K K$ decay and diffusion processes (e.g. $\phi \bar K \to \bar K$), since Bose enhancement is more effective on the light modes of the system \cite{Cabrera:2013iya}.

Apart from the temperature and density dependence of the scattering amplitudes and the corresponding behavior of the hyperons in matter, an additional output of our model is the momentum dependence of nuclear optical potentials. This is important in order to have a comprehensive description of medium effects on all the hadrons involved in strangeness production near threshold. Moreover, hadronic medium effects can only be implemented by means of the quasi-particle prescription in a certain class of transport models such as the Isospin Quantum Molecular Dynamics approach (see \cite{Hartnack:2011cn} and references therein), where the physical states are implemented according to on-shell kinematics.

Particularly for the $\Lambda$, $\Sigma$ and $\Sigma^*$ hyperons, which essentially keep a quasi-particle nature in the medium (with some caveats with the $\Sigma^*(1385)$, largely broadened in the medium), we can determine the hyperon optical potential by analyzing the momentum evolution of the resonance pole in the scattering amplitudes. On one hand, the real part of the optical potential can be obtained by subtracting the free hyperon dispersion relation from the in-medium quasi-particle energy, $\epsilon_{Y}(P)$,
\begin{equation}
{\rm Re} \ U_{Y}(P) = \epsilon_{Y}(P) - \sqrt{M_Y^2+\vec{P}^2} \ .
\end{equation}
On the other hand, a suitable combination of the amplitude residue at the resonance pole and the imaginary part of the amplitude evaluated at the quasi-particle energy allows to calculate the acquired width in the medium (total width including the vacuum one in the case of the $\Sigma^*$):
\begin{equation}
{\rm Im}\ U_{Y}(P) = {\rm Im} \ T_{ij}(\epsilon_{Y}(P)) / m
\end{equation}
with $m$ being the slope of ${\rm Re}\ T_{ij}$ at the resonance pole. We note that this definition is equivalent to the more general definition of the hyperon self-energy, $\Sigma_Y$, as elaborated in Ref.~\cite{Kaskulov:2005uw}. Our $p$-wave amplitudes are driven by the spectral function (or, equivalently, the propagator) of the hyperons, which reads
\begin{equation}
\label{SF}
S_{Y}(P^0,\vec{P}) = 
-\frac{1}{\pi} 
\frac{M_Y}{E_{Y}(\vec{P})} 
\frac{{\rm Im} \,\Sigma_{Y}(P^0,\vec{P})}
{[P^0-E_Y(\vec{P})-{\rm Re}\, \Sigma_{Y}(P^0,\vec{P})]^2
+[{\rm Im} \, \Sigma_{Y}(P^0,\vec{P})]^2} \ .
\end{equation}
The optical potential defined above corresponds to the hyperon self-energy evaluated at the quasi-particle energy for a given momentum.

Following this method we provide in Fig.~\ref{fig:UY} the momentum-dependent nuclear optical potentials for the $\Lambda(1115)$, $\Sigma(1195)$ and $\Sigma^*(1385)$ hyperons for several densities and temperatures. The evolution of the optical potentials with nuclear density and temperature can be easily traced back to the shape and position of the hyperon peaks in the isospin amplitudes previously discussed.

The real (imaginary) part of the optical potential is displayed in the upper (lower) panel of Fig.~\ref{fig:UY}.
At normal nuclear matter density, the $\Lambda$, $\Sigma$ and $\Sigma^*$ acquire attractive potentials of -$48$, -$40$ and -$10$~MeV, respectively, at rest in the nuclear matter rest frame. At densities beyond $\rho_0$, the attraction on the $\Lambda$ and $\Sigma$ is enhanced whereas the potential for the $\Sigma^*$ turns from attractive to repulsive between $\rho_0$ and $2\rho_0$. The momentum dependence is rather smooth in all three cases: the potentials monotonically increase (thus the attraction being reduced). For the $\Sigma^*$, which experiences a rather small binding, the potential turns from attractive to repulsive at about $500$~MeV$/c$ momentum.
The temperature mildly reduces the size of the potential for the $\Lambda$ and $\Sigma$ hyperons, in line with the input baryonic binding potentials implemented in the intermediate hyperon propagators. For the $\Sigma^*$ the optical potential is tied to medium effects on the main decay channels already existing in vacuum, as already discussed,
where typically large cancellations between real parts in the self-energy (from different channel contributions) lead to only moderate shifts of the resonance mass \cite{Kaskulov:2005uw}.
We obtain in this case that the real part of the $\Sigma^*$ potential is slightly larger in magnitude (more attractive) at $\rho=\rho_0$ and $T=100$~MeV as compared to the zero temperature case.

The imaginary part of the optical potential is due to the opening of in-medium decay or absorption channels involving the interactions with nucleons. Both the $\Lambda$ and the $\Sigma$ evolve from being stable states in vacuum to having relatively small decay widths, below 20~MeV for the range of density, temperature and momentum studied here. We recall again that the $\Lambda$ can only decay through the excitation of $NN^{-1}$ components, which require a finite hyperon momentum at zero temperature. Whereas one may expect larger widths for $\Lambda$ and $\Sigma$ at $\rho=2\rho_0$ as compared to $\rho=\rho_0$, the enhancement of the available states with density is compensated by the shift in mass of the hyperons, leading to a small reduction in the width for both $J^P=1/2^+$ baryons at low momentum.
We recall that for large densities, self-consistency for the hyperon single-particle potential might be required.
Moreover, the value of the hyperon width at the quasi-particle energy (as obtained from ${\rm Im}\,U_Y$) may differ from the one developed by the (off-shell) spectral function, particularly when the energy dependence of the self-energy is substantial, as is the case for $p$-wave interactions.
The density evolution of the width for the $\Sigma^*$ essentially reflects the enhancement of in-medium phase space of its decay channels. At $\rho=\rho_0$ and $T=100$~MeV, however, we find a small decrease of the width with respect to the $T=0$ case, which is traced back to the baryons in the final state, becoming heavier with increasing temperature.

\section{In-medium transition probabilities and cross sections}   
\label{sec:cross-sec}

The dynamics of the $\bar KN$ system and its related coupled channels in the hot and dense medium is encoded in the $S=-1$ meson-baryon scattering amplitudes. With the focus on the implementation, in transport simulations, of strangeness dynamics in heavy-ion collisions at FAIR conditions we present our analysis in terms of transition probabilities and cross sections for different binary reactions. These results are complementary to the $\bar K$ spectral functions and nuclear optical potentials provided in Ref.~ \cite{Tolos:2008di}
and, altogether, permit a systematic accounting of medium effects in the $S=-1$ sector, not only within the relevant binary reactions but also regarding the production and propagation of light strange hadrons.

In general the calculation of dynamical quantities in transport theory will require an appropriate folding of reaction rates or transition probabilities with the spectral functions of the particles in the initial and final states. Such is the case of the model in \cite{Cassing:2003vz}, which is based on a gradient expansion of the Kadanoff-Baym equation and accounts for the transport of off-shell particles.
The transition probability for a given reaction, ${\cal P}(s)$, is determined as the angular integrated average squared amplitude (including all partial waves) and can be defined fully off-shell as a function of the total energy $P^0$ and momentum $\vec{P}$ of the meson-baryon pair. For the process $i \to j$ (where $i,j$ denote meson-baryon channels) one has
\begin{equation}
\label{eq:P-of-s}
{\cal P}_{ij}(P^0,\vec{P};\rho,T) = \int_{-1}^{+1} du
\lbrace
| f^{(s)}_{ij} + (2f^+_{ij}+f^-_{ij}) \, u |^2
+
| f^+_{ij} - f^-_{ij} |^2 \,(1-u^2)
\rbrace \ ,
\end{equation}
where $f^{(s)} = T^{L=0}$, $f^{\pm}$ is given in Eqs.~(\ref{fg}-\ref{g1}) in terms of suitable combinations of spin-flip and spin-non-flip $p$-wave amplitudes, and  $\theta$ is the scattering angle in the c.m. frame of the meson-baryon pair.

We note that, modulo kinematical factors related to flux of the incoming and outgoing particles, the former expression recalls that of the total cross section for a binary process in vacuum. The definition of an in-medium cross section, however, is more complex and requires both the knowledge of the pertinent scattering amplitudes at finite temperature and density, and a suitable generalization of the corresponding flux factors. Taking into account that the hadrons in the initial and final states do need not be on the mass shell (as they could develop a broad spectral function in the medium), there is not a unique and simple way to implement such definition in the medium and requires the choice of an on-shell reduction scheme \cite{Cassing:2003vz,Hartnack:2011cn}.
Still, medium effects for strange reactions are best implemented in terms of in-medium cross sections (and pertinent nuclear optical potential) in transport models which rely on the narrow test quasiparticle approach \cite{Hartnack:2011cn}, and thus we deem pertinent to provide also in-medium cross sections from our meson-baryon scattering amplitudes, which we discuss in the second half of this section.
The differential cross section for the process $i\to j$ (where $i,j$ denote meson-baryon channels) reads
\begin{equation}
\label{eq:diff-cross-sec}
\frac{d\sigma_{ij}}{d\Omega} = \frac{1}{16 \pi^2} \frac{M_i M_j}{s} \frac{\tilde{q}_j}{\tilde{q}_i}
\lbrace
| f^{(s)}_{ij} + (2f^+_{ij}+f^-_{ij}) \cos\theta |^2
+
| f^+_{ij} - f^-_{ij} |^2 \sin^2\theta
\rbrace \ ,
\end{equation}
with $\tilde{q}_i$ the c.m. three-momentum of meson-baryon pair $i$. The total cross section follows as
\begin{equation}
 \sigma_{\rm tot} = \int d\Omega \frac{d\sigma_{ij}}{d\Omega} = 2\pi\int_{-1}^{1} du \, \frac{d\sigma_{ij}}{d\Omega}(u) \ ,
\end{equation}
 with $u\equiv\cos\theta$.

We discuss next the transition probability for several $K^- p$ reactions. From here on we shall denote these rates as ${\cal P}(s)$, keeping in mind that in our model they actually depend separately on the total energy $P^0$ and momentum $\vec{P}$ of the meson-baryon pair. In the following discussion, we present selected results for ${\cal P}(s)$ at zero total momentum, $P=0$, as a function of $s^{1/2}=P^0$ for several nuclear densities and temperatures.

\begin{figure}[t]
\begin{center}
\includegraphics[height=9cm]{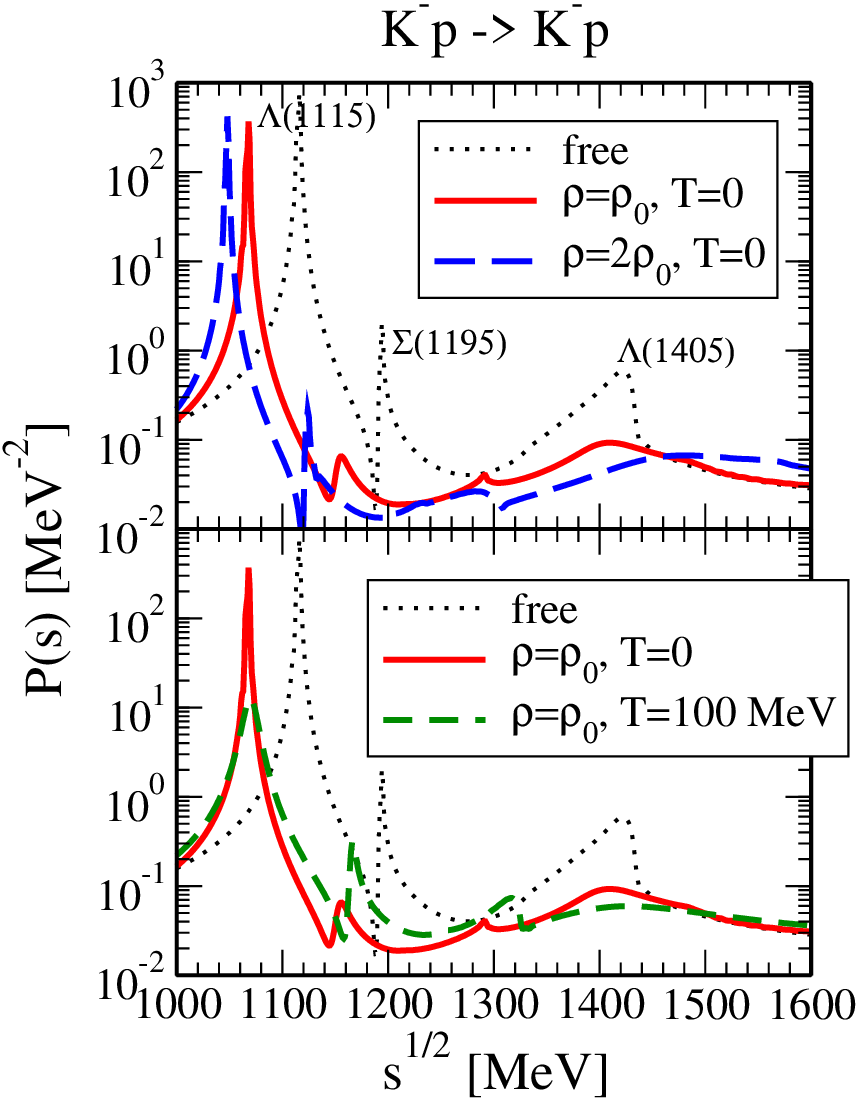}
\includegraphics[height=9cm]{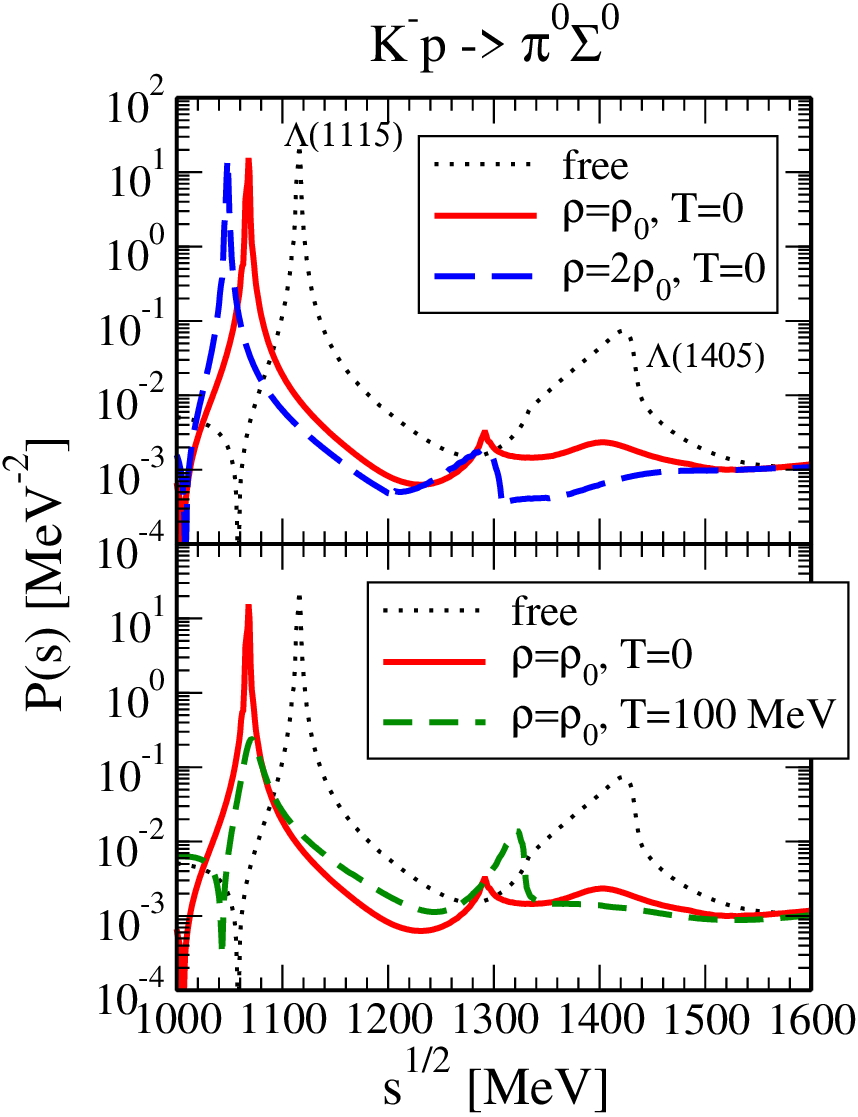}
\caption{In-medium transition probability ${\cal P}(s)$ at zero total three-momentum for the elastic $K^-p$ (left) and the inelastic $K^- p \to \pi^0 \Sigma^0$ (right) reactions. The peaks associated to the $\Lambda(1115)$, $\Sigma(1195)$ and $\Lambda(1405)$ resonances are clearly visible in the vacuum case.}
\label{fig:Ptrans-Kmp}
\end{center}
\end{figure}

In Fig.~\ref{fig:Ptrans-Kmp} we depict the transition probability for the $K^-p$ elastic reaction and the $K^-p \to \pi^0 \Sigma^0$ strangeness exchange reaction.
The $K^-p$ state is an admixture of $I=0,1$ and therefore the two isoscalar $\Lambda$ resonances and the isovector $\Sigma(1195)$ show up according to the results discussed in Sec.~\ref{sec:amplitudes}.
The $\Sigma^*(1385)$ couples weakly to the $\bar K N$ system and cannot be resolved in the $K^-p$ elastic case.
The $K^-p \to \pi^0 \Sigma^0$ reaction selects the $I=0$ component of the $\bar K N$ amplitude and consequently only the isoscalar hyperons are present in the right panel of Fig.~\ref{fig:Ptrans-Kmp}.
The resonance profiles exhibit the temperature and density evolution as discussed for the amplitudes in Sec.~\ref{sec:amplitudes}. The structure of the $\Lambda(1405)$ is practically washed out and only some remnants are visible at normal matter density. The effect of temperature is particularly appreciable as a broadening of the $p$-wave resonances as compared to the vacuum case (recall that in vacuum the $\Lambda$ and $\Sigma$ are stable and their apparent width is a numerical artifact).

\begin{figure}[t]
\begin{center}
\includegraphics[height=9cm]{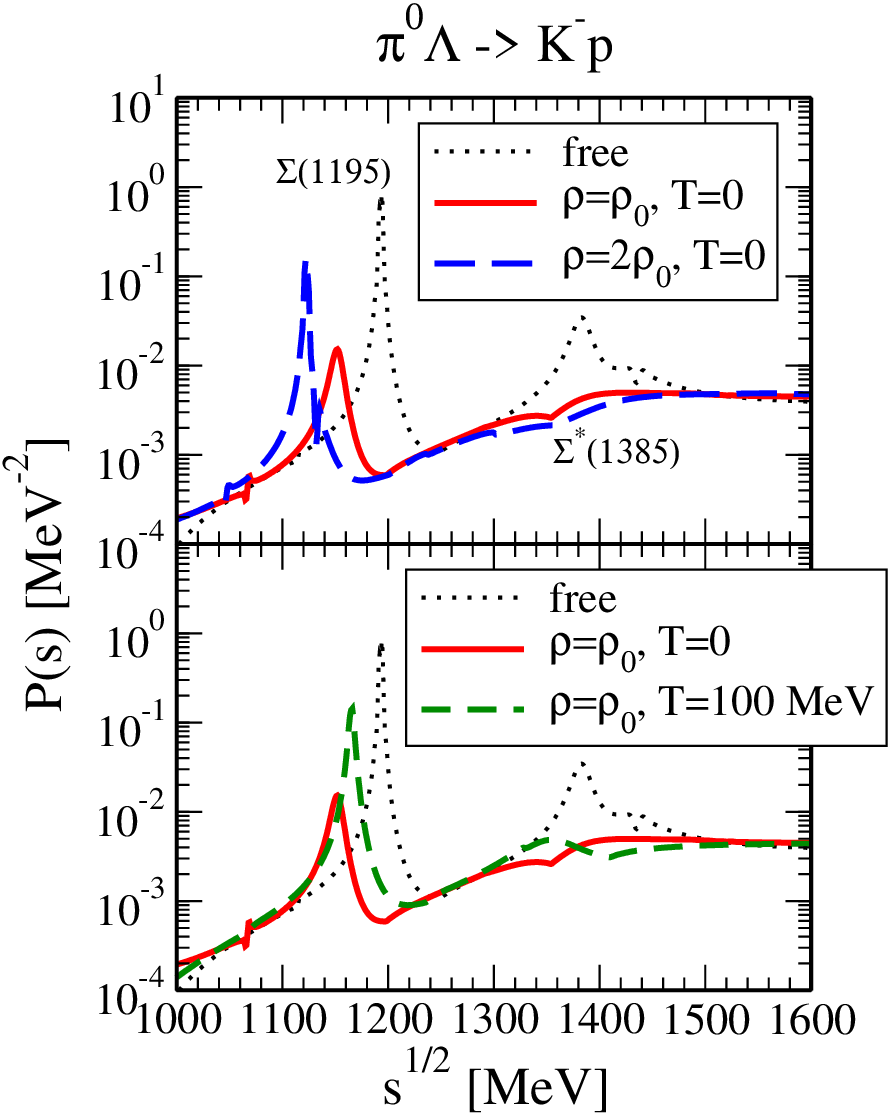}
\includegraphics[height=9cm]{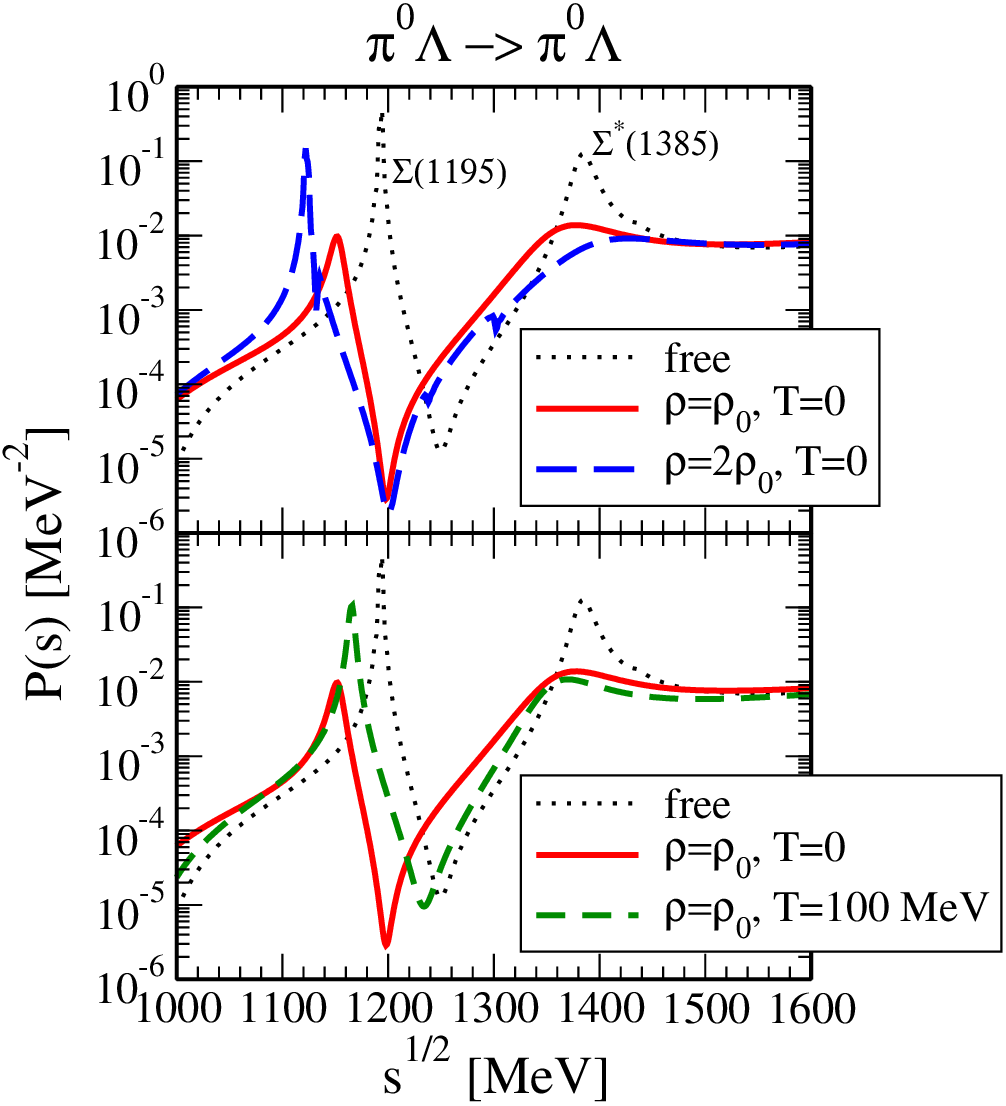}
\caption{Same as in Fig.~\ref{fig:Ptrans-Kmp} for the inelastic $K^-p\to \pi^0 \Lambda$ (left) and  the $\pi^0\Lambda\to \pi^0\Lambda$ (right) $I=1$ reactions. The peaks associated to the $\Sigma(1195)$ and the $\Sigma^*(1385)$ resonances are clearly visible in the vacuum case.}
\label{fig:Ptrans-I1}
\end{center}
\end{figure}

Next we show in Fig.~\ref{fig:Ptrans-I1} the transition probability for the pure isovector processes $\pi^0\Lambda \to K^- p$ and $\pi^0 \Lambda \to \pi^0 \Lambda$, where in this case only the $\Sigma$ resonances populate the spectrum.
The $\Sigma^*(1385)$ couples strongly to $\pi \Lambda$ and more weakly to $\pi\Sigma$ and $\bar K N$ states. The latter channel is actually closed in vacuum. However, at finite nuclear density the $\bar K N$ threshold is lowered because of the attractive potentials acting on the meson and the baryon. Because of the opening of this channel (and related in-medium processes accounted for in the $\bar K$ self-energy) one observes that the $\Sigma^*$ shape is distorted by the in-medium $\bar K N$ threshold and its signal practically disappears in the $\pi^0 \Lambda \to K^- p$ reaction. In the diagonal process $\pi^0 \Lambda \to  \pi^0 \Lambda$, one can appreciate the large in-medium width of the $\Sigma^*$ induced by the one- and two-body mechanisms incorporated through the dressing of pions and kaons, as well as the small changes in the position of the resonance.

\begin{figure}[t]
\begin{center}
\includegraphics[height=8cm]{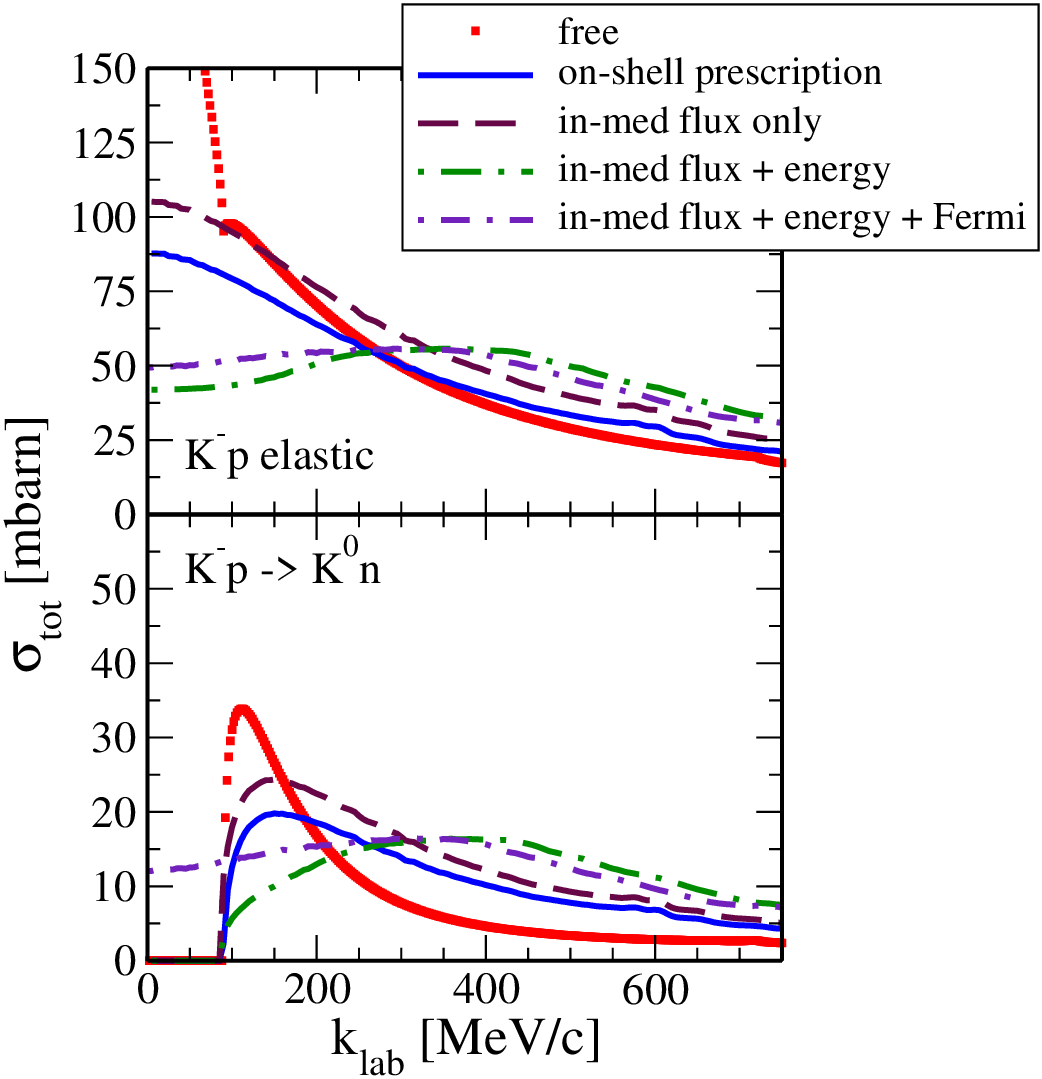}
\caption{Total $K^- p \to K^- p$ and $K^- p \to \bar K^0 n$ cross sections with in-medium amplitudes.
The different curves show the cross section if medium effects on the initial/final state two-particle flux, the meson/baryon energies and the Fermi motion are included (see text).}
\label{fig:cross-sec-med-effects}
\end{center}
\end{figure}
\begin{figure}[t]
\begin{center}
\includegraphics[height=6.5cm]{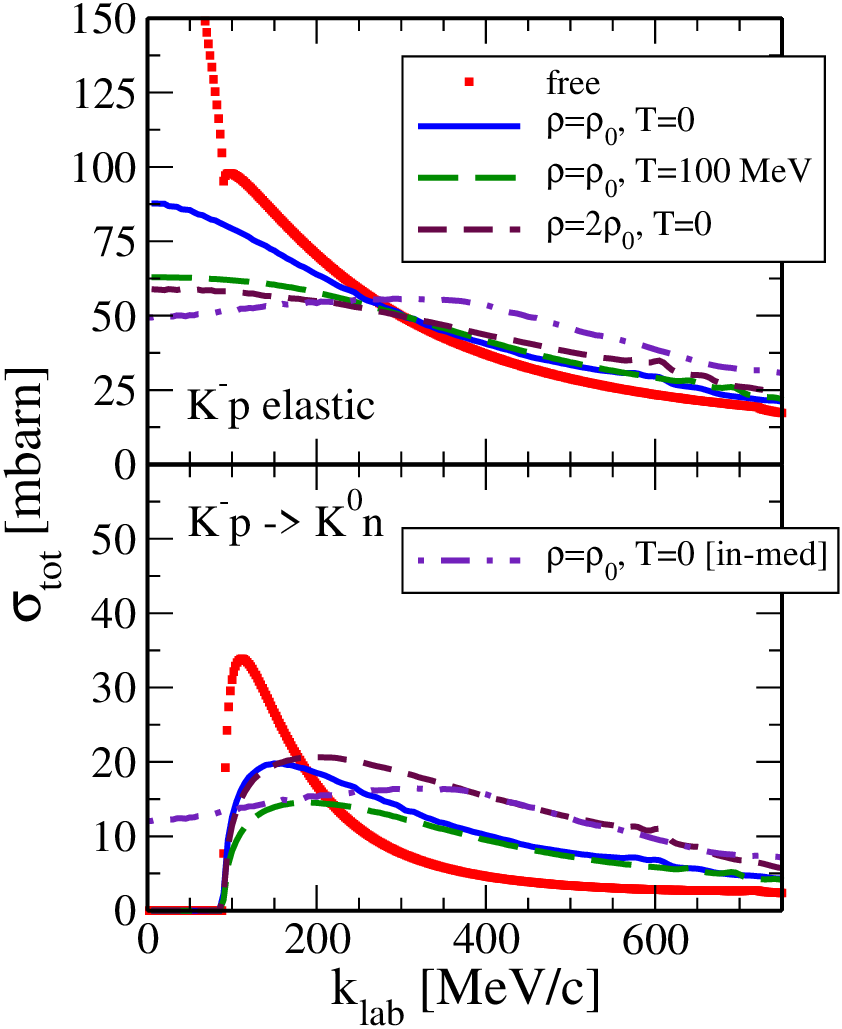}
\includegraphics[height=6.5cm]{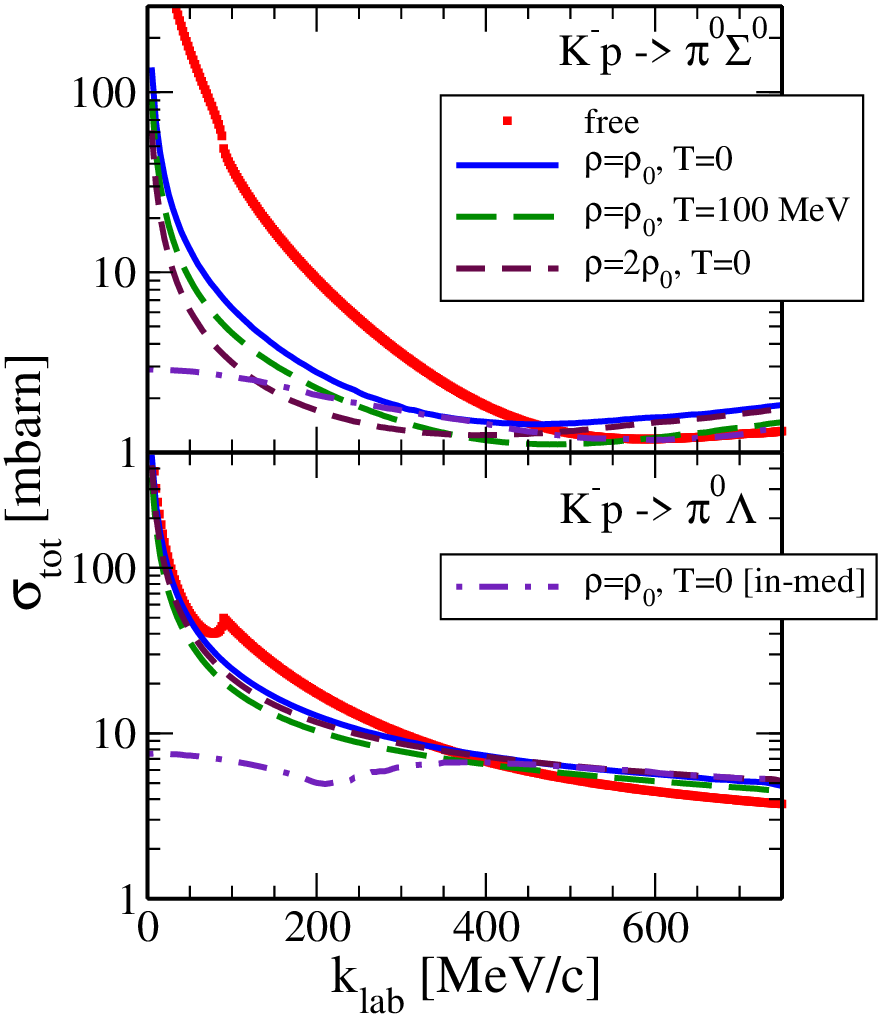}
\includegraphics[height=6.5cm]{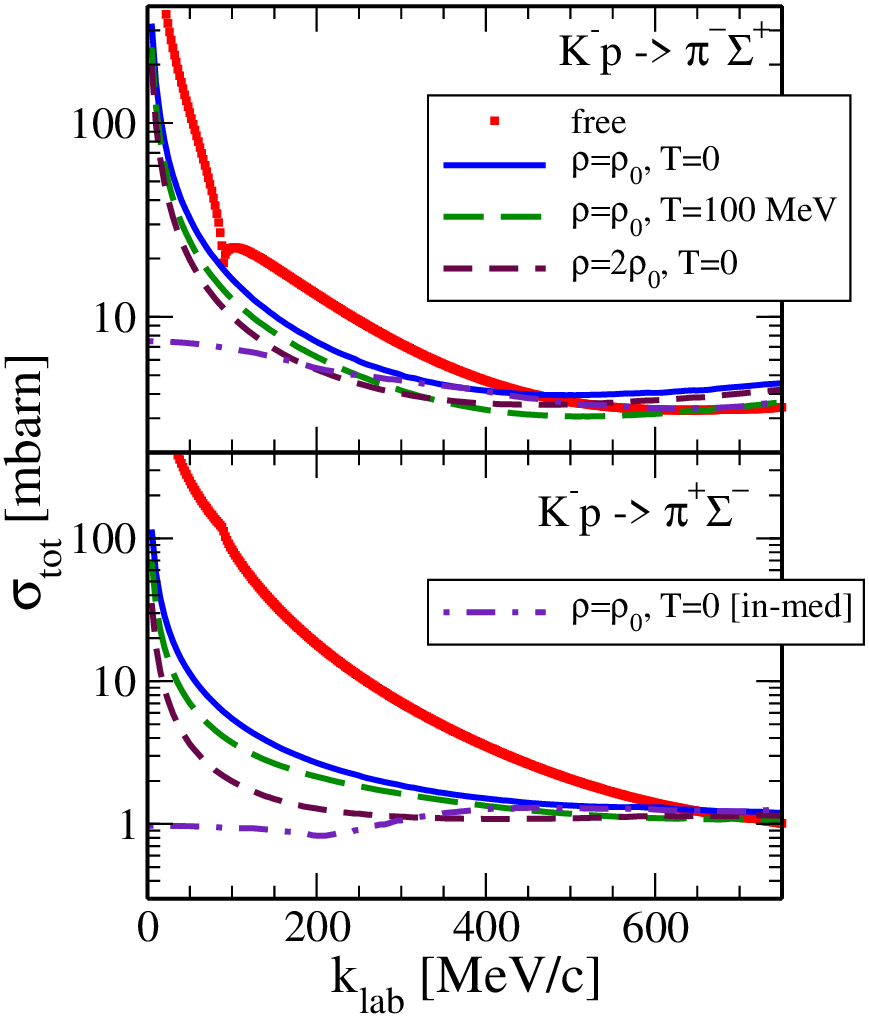}
\caption{Total $K^- p$ cross sections with in-medium amplitudes including $s$- and $p$-waves for several coupled channels.}
\label{fig:cross-sec-med}
\end{center}
\end{figure}

In the following we present results for the in-medium cross sections of the $K^- p$ elastic and inelastic binary reactions, which we compare with the vacuum ones.
The simplest way to estimate the in-medium cross section for these processes is to replace in Eq.~(\ref{eq:diff-cross-sec}) the amplitudes in vacuum by their in-medium counterparts.
The results are shown in Fig.~\ref{fig:cross-sec-med-effects}, with a solid line (on-shell prescription), and in Fig.~\ref{fig:cross-sec-med} for the elastic case and several inelastic channels involving strangeness exchange (thick lines) at different temperature and density. As a common feature we observe that the rapid fall of the cross section close to threshold is softened and the strength is distributed over a wide range of energies, as expected from the melting of the $\Lambda (1405)$ resonance in matter at finite temperature. Typically, the in-medium cross section overshoots the vacuum one at finite momentum (this happens, e.g., in elastic $K^-p$ for $K^-$ momenta in the lab $\gtrsim 300$~MeV$/c$).

As discussed at the beginning of this section, some caveats emerge from the definition of the in-medium cross section in the on-shell prescription.
First, in vacuum the incident kaon momentum in the lab frame determines the total energy and momentum of the $K^-p$ pair and thus the center-of-mass energy, $\sqrt{s}$. Then the evaluation of the scattering amplitude is straightforward since it only depends on $s$ (in general on the invariants $s$, $t$ and $u$) in vacuum.
However, in the nuclear medium, the Lorentz covariance is broken and the (off-shell) scattering amplitudes depend explicitly on $P^0$ and $\vec{P}$. In the nuclear matter rest frame and neglecting the Fermi motion of the initial nucleon one has $\vec{P}=\vec{q}$ and $P^0=q^0+M_N$, where $q^0$ is the off-shell energy of the incoming antikaon. Since there is not a unique relation between $q^0$ and $\vec{q}$ the probability for this reaction to occur should be folded with the spectral function of the antikaon. Otherwise the information about the in-medium properties of the strange mesons, encoded in the meson self-energies, is not taken into account. Note, for instance, that the $\bar K$ is attracted by -$45$~MeV at $\rho=\rho_0$ and therefore the total energy of the $\bar K N$ pair at a given lab momentum is lower than within the free on-shell prescription ($P^0=\sqrt{m_K^2+\vec{q}\,^2}+M_N$), thus giving access to the energy region below the nominal $\bar K N$ threshold in vacuum.
An educated estimation of these effects on the effective cross section for $\bar K N$ scattering in the nuclear medium can be addressed as follows. An incident antikaon with momentum $\vec{q}$ in the nuclear matter rest frame will have an energy distribution which we can approximate by the narrow quasi-particle energy, $q^0\simeq \omega(q)+U_{\bar K}(q)$ with $U_{\bar K}=\Pi_{\bar K}/2m_K$ the $\bar K$ nuclear optical potential and $\omega(q)=\sqrt{m_K^2+\vec{q}\,^2}$ (the former provides a good approximation to the exact dispersion relation for the present purpose).
Then the total energy and momentum of the meson-baryon pair is given by
\begin{eqnarray}
\label{eq:P0Pmed}
P^0 &=& \omega(q) + U_{\bar K}(q) + M_N^* + \Sigma_N^v  \ , \nonumber \\
P &=& q       \ ,
\end{eqnarray}
where the nucleon energy is also modified by the corresponding scalar ($\Sigma_N^s$) and vector ($\Sigma_N^v$) mean field potentials ($M_N^*$ contains the scalar part, $M_N^*=M_N-\Sigma_N^s$), and where we have assumed for simplicity that the initial nucleon is at rest. It is clear from the equations above that the effective squared invariant energy $s^*=(P^0)^2-P^2$ is lower than its value in vacuum due to the attractive potentials acting on both the $\bar K$ and the nucleon.
Of course in the nuclear matter rest frame the initial nucleon is not at rest but vibrates with Fermi motion. In order to estimate the effect of the Fermi motion of the initial nucleon we obtain the angular average over the nucleon distribution, which amounts to modifying Eq.~(\ref{eq:P0Pmed}) as follows,
\begin{equation}
\label{eq:Fermiav}
M_N^* \to \sqrt{(M_N^*)^2+\frac{3}{5}p_F^2(\rho)} \quad , \quad
P = q \to P = \sqrt{q^2+\frac{3}{5}p_F^2(\rho)} \, ,
\end{equation}
with $p_F(\rho)$ such that $\rho=2p_F^3/3\pi^2$.
Similarly, the c.m. incoming and outgoing momenta $\tilde{q}$ and $\tilde{q}'$ in Eq.~(\ref{eq:diff-cross-sec}) are modified to take into account the meson-baryon binding.
The effect of these corrections is analyzed in Fig.~\ref{fig:cross-sec-med-effects} for the $K^- p$ elastic and $K^- p \to \bar K^0 n$ reactions at normal nuclear matter density and zero temperature.
In these particular channels the binding on the initial and final states is the same and thus the modification of the kinematical factors in the cross section simply reflects the reduction of $s^*$ with respect to the vacuum case, which induces a moderate increase in the cross section (cf.~compare curves labelled with ``on-shell prescription" and ``in-medium flux only'', where in the latter only the kinematical factors $\frac{1}{s} \,\tilde{q}_j/\tilde{q}_i$ are modified). When the amplitudes are evaluated at total energy and momentum accounting for the nuclear potentials, cf.~Eq.~(\ref{eq:P0Pmed}) and curves labelled with "in-med flux+energy", we find that the strength is substantially redistributed to higher lab momenta. This results from the fact that, due to the nuclear binding in the initial state, the energy required to excite the $\Lambda(1405)$ can only be reached at a finite momentum of the incident antikaon. Finally, the Fermi motion of the initial nucleon (cf.~curves "in-med flux+energy+Fermi'') 
only enhances moderately the cross section at small incident momentum in the elastic channel, whereas for $K^- p\to\bar K^0 n$ the threshold is shifted to lower energies and this channel is open even  below the $\bar K^0 n$ vacuum threshold (with relative momentum of the meson-baryon pair in the c.m. frame of about $100$~MeV$/c$).
We refer to the dash-dotted lines in Fig.~\ref{fig:cross-sec-med} for an estimation of these modifications in the inelastic $K^-p$ reactions.

\section{Summary, conclusions and outlook}
\label{sec:Conclusion}
We have extended our model for $S=-1$ meson-baryon interaction in hot and dense nuclear matter by incorporating the $p$-wave amplitudes within the unitarized self-consistent scheme that was already built in for the $s$-wave. This has allowed us to compute scattering amplitudes for binary kaon-nucleon reactions in different diagonal and off-diagonal coupled channels, for isospin $I=0,1$ and total spin $J=1/2,3/2$.

The isoscalar, $s$-wave $\bar K N$ amplitude is dominated by the excitation of the $\Lambda(1405)$ right below threshold, which acquires  its physical width dominantly from the decay into $\pi \Sigma$ states. When the nuclear medium is switched on, the resonance  is practically washed out and its strength spread out over energy, as a consequence of the in-medium decay mechanisms incorporated through the self-consistent dressing of mesons (e.g. $\Lambda(1405) \to \pi (YN^{-1}) N, \pi (NN^{-1})\Sigma, \pi (\Delta N^{-1})\Sigma$).

The $p$-wave amplitude reflects the excitation of the $\Lambda$ and $\Sigma$ hyperons (in isospin 0 and 1, respectively) in the spin $1/2$ channel, and the $\Sigma^*(1385)$ with spin $3/2$. At finite nuclear densities both the $\Lambda$ and the $\Sigma$ experience an attractive potential of roughly -50 and -40~MeV at normal matter density and zero temperature, consistently with the input mean-field of the $\sigma-\omega$ model employed to account for medium effects in the baryon propagators of intermediate meson-baryon states. Both hyperons acquire a finite decay width, reflecting the probability to be absorbed by the nuclear medium or have quasi-elastic scattering processes at finite density and temperature. The $\Sigma^*$ develops a much smaller attractive potential of about -10~MeV at $\rho=\rho_0$ and zero temperature which turns even to a small repulsion for increasing densities. Its decay width is notably enhanced by a factor three at normal density mostly due to the dressing of pions, which opens new 
absorption channels such as $\Sigma^* \to \pi (NN^{-1}) \Lambda, \pi (NN^{-1}) \Sigma$ and similarly with the pion coupling to $\Delta N^{-1}$ excitations. The effect of the temperature in this case is moderate due to the important phase space already available  at zero temperature.

An additional output of the model, which can be accessed from the $p$-wave amplitudes, is the momentum, density and temperature dependent optical potential of the $\Lambda$, $\Sigma$ and $\Sigma^*$. In all cases we have observed a smooth behavior with momentum up to 500 MeV/c.

We have exploited the novel features in our model in order to calculate the in-medium total cross section for the $K^-p$ elastic and inelastic reactions and compared our results with the vacuum case. These cross sections, dominated by the $s$-wave interaction, are particularly smoothened at low incident momenta and some strength is extended to energies below threshold due to the effectively smaller mass of antikaons in the dense medium. As a consequence of the melting of the $\Lambda(1405)$ the cross sections fall off more slowly and eventually remain larger than the vacuum ones with increasing energy.

Our in-medium scattering amplitudes have also been used to generate off-shell transition rates for binary reactions involving strange mesons, such as $\bar K N \to \bar K N$ and $\pi\Lambda \to \bar K N$, of crucial importance to understand strangeness production mechanisms in heavy-ion collisions.
The implementation of this dynamical information together with the spectral functions of $\bar K$ in a suitable off-shell transport model along the line of Ref.~\cite{Cassing:2003vz} is on-going and will be reported elsewhere \cite{preparation}. Also,  results on strange vector mesons in matter have been recently reported ($\bar K^*$ \cite{Tolos:2010fq} and $K^*$ \cite{Ilner:2013ksa}) and should be implemented in the transport models in order to have a unified scheme for strangeness production and dynamics in heavy-ion collisions.

\section*{Acknowledgements}
We acknowledge fruitful discussions with Wolfgang Cassing, Eulogio Oset and \`Angels Ramos. This work has been supported by the Helmholtz International Center for FAIR within the framework of the LOEWE program. We also acknowledge support from Grants No. FPA2010-16963 (Ministerio de Ciencia e Innovaci\'on), No. FP7-PEOPLE-2011-CIG under Contract No. PCIG09-GA-2011-291679 and the European Community-Research Infrastructure Integrating Activity Study of Strongly Interacting Matter (acronym HadronPhysics3, Grant Agreement n. 283286) under the Seventh Framework Programme of EU. D.C. acknowledges support from the BMBF (Germany) under project no.~05P12RFFCQ.  L.T. acknowledges support from the Ram\'on y Cajal
Research Programme (Ministerio de Ciencia e Innovaci\'on).

\appendix

\section{Antikaon self-energy from $T_{{\bar K} N}$}
\label{app-swave}

We derive in this section a general expression for the antikaon
self-energy from the effective in-medium $\bar K N$ scattering amplitude at
finite temperature. Let us denote by $T_{{\bar K} N}$ the isospin averaged
antikaon nucleon scattering amplitude. The argument is valid for both $s$ and $p$-waves and thus we omit the $L$ index in the notation.
Following the Feynman rules in the ITF, 
the $\bar K$ self-energy reads
\begin{equation}
\label{s-wave-ITF}
\Pi_{{\bar K} N} (\omega_n,\vec{q}; T) = T\, \sum_{m=-\infty}^{\infty}
\int \frac{d^3p}{(2\pi)^3} \, \frac{1}{{\rm i} W_m - E_N (\vec{p}\,)}
\, T_{{\bar K} N} (\omega_n + W_m , \vec{P} ; T)
\ ,
\end{equation}
where $\omega_n$ and $W_m$ are bosonic and fermionic Matsubara frequencies,
respectively, ${\rm i} \omega_n = {\rm i} 2n\pi T$ and
${\rm i} W_m = {\rm i} (2m+1)\pi T + \mu_B$.
The sum over the
index $m$ cannot be solved unless we know exactly how $T_{{\bar K} N}$ depends on $m$. We skip this difficulty by invoking a spectral
representation for the $T$-matrix, i.e.
\begin{eqnarray}
\label{s-wave-ITF-2}
\Pi_{{\bar K} N} (\omega_n,\vec{q}; T)
&=&
-T\, \sum_{m=-\infty}^{\infty}
\int \frac{d^3p}{(2\pi)^3} \,
\frac{1}{\pi}\int_{-\infty}^{\infty} d\Omega \,
\frac{{\rm Im}\,T_{{\bar K} N}(\Omega,\vec{P};T)}
{[{\rm i} W_m - E_N (\vec{p}\,)] [{\rm i} \omega_n + {\rm i} W_m - \Omega]}
\nonumber \\
&=&
- \int \frac{d^3p}{(2\pi)^3} \,
\frac{1}{\pi}\int_{-\infty}^{\infty} d\Omega \,
\frac{{\rm Im}\,T_{{\bar K} N}(\Omega,\vec{P};T)}
{{\rm i} \omega_n - \Omega + E_N (\vec{p}\,)}
\, [n_N(\vec{p},T) - n(\Omega,T)] \, ,
\nonumber \\
\end{eqnarray}
with $n(\Omega,T) = [e^{(\Omega-\mu_B)/T}+1]^{-1}$ here.
The former result, after
continuation into the real energy axis (${\rm i}\omega_n \to q_0+{\rm
i}\varepsilon$), provides the finite-temperature antikaon self-energy evaluated from the antikaon
nucleon scattering amplitude. Note that it includes a Pauli blocking correction
term, $n(\Omega,T)$, convoluted with the spectral strength from the imaginary
part of the $T$-matrix. At the region in which the principal value of the
spectral integration gets its major contribution, $\Omega \approx q_0 +
E_N(\vec{p}\,)$, the fermion distribution $n(\Omega,T)$ behaves as a slowly
varying function and thus we can
approximate this term by a constant, namely, $n(\Omega,T)\simeq n(q_0 +
E_N(\vec{p}\,),T)$ and take it out of the integral. The dispersion
integral over $\Omega$ then recovers the whole amplitude $T_{{\bar K} N}$ and
the self-energy can be approximated by:
\begin{equation}
\label{s-wave-ITF-3}
\Pi_{{\bar K} N} (q_0+{\rm i}\varepsilon,\vec{q}; T) =
\int \frac{d^3p}{(2\pi)^3} \,
T_{{\bar K} N} (q_0 + E_N(\vec{p}\,),\vec{P};T) \,
[n_N(\vec{p},T) - n(q_0 + E_N(\vec{p}\,),T)]
\ .
\end{equation}
Note that this procedure is exact for the imaginary part.
Eqs.~(\ref{eq:selfd},\ref{eq:selfd-v2}) follow from the former result by neglecting the second term in the brackets. The latter is a sensible approximation for the $s$-wave amplitude, which peaks
around the $\Lambda(1405)$ resonance, and thus one expects this correction to be
small with respect to that on the nucleon. For the $p$-wave self-energy, the two terms in the brackets guarantee that the crossing property of the (retarded) self-energy, ${\rm
Im} \, \Pi_{\bar K}^p (-q_0,\vec{q};T) = - {\rm Im} \, \Pi_K^p
(q_0,\vec{q};T)$, is satisfied. As a consistency test, we find that the $p$-wave self-energy derived in \cite{Tolos:2008di} in terms of $YN^{-1}$ Lindhard functions and $\bar KNY$ amplitudes at leading order is recovered from Eq.~(\ref{s-wave-ITF-3}) if replacing the $T$-matrix by the tree-level (hyperon pole) amplitude in Eqs.~(\ref{fg}-\ref{g1}).

  \end{document}